\documentclass[a4paper,12pt]{article}
\pdfoutput=1 

\usepackage{jheppub} 
                     
\usepackage{bbold}                     
\usepackage{lmodern} 
\usepackage[T1]{fontenc} 
\usepackage[utf8]{inputenc} 

\newcommand{\g}{\gamma}
\newcommand{\ta}{\theta}
\newcommand{\la}{\lambda}
\newcommand{\pd}{\partial}
\newcommand{\al}{\alpha}
\newcommand{\bt}{\beta}

\newcommand{\ep}{\epsilon}
\newcommand{\sg}{\sigma}
\newcommand{\dt}{\delta}

\newcommand{\La}{\Lambda}
\newcommand{\nn}{\nonumber}
\newcommand{\alh}{\widehat{\alpha}}

\newcommand{\tah}{\widehat{\theta}}
\newcommand{\lh}{\widehat{\lambda}}
\newcommand{\wh}{\widehat{w}}

\newcommand{\ab}{\underline{a}}
\newcommand{\bb}{\underline{b}}
\newcommand{\cb}{\underline{c}}
\newcommand{\db}{\underline{d}}

\newcommand{\wb}{\overline{w}}
\newcommand{\lb}{\overline{\lambda}}

\newcommand{\pb}{\overline{\partial}}

\newcommand{\zb}{\overline{z}}

\usepackage[compat=1.1.0]{tikz-feynman}
\usepackage{subcaption}
\usepackage{esvect}

\title{\boldmath Vertex operators for the superstring with manifest $d=6$ $\mathcal{N}=1$ supersymmetry}

\author[]{Cassiano A. Daniel}

\affiliation[]{ICTP South American Institute for Fundamental Research
\\Instituto de F\'isica Te\'orica, Universidade Estadual Paulista, \\
Rua Dr. Bento Teobaldo Ferraz 271, 01140-070, S\~ao Paulo - SP, Brasil}

\emailAdd{c.daniel@unesp.br}

\abstract{By extending the six-dimensional hybrid formalism for the superstring to include $d=6$ $\mathcal{N}=1$ superspace variables along with unconstrained bosonic ghost fields, we construct a manifestly spacetime supersymmetric vertex operator $U$. We demonstrate that the BRST invariance of $U$ implies the $d=6$ $\mathcal{N}=1$ SYM equations of motion in superspace. Furthermore, we show that spacetime supersymmetric scattering amplitudes can be computed in a manner analogous to that of the non-minimal pure spinor formalism.

}

\begin{document}
\maketitle
\flushbottom

\section{Introduction}

Alternative formulations of the superstring with manifest spacetime supersymmetry (SUSY), such as the hybrid \cite{Berkovits:1996bf} \cite{Berkovits:1994vy} and pure spinor \cite{Berkovits:2000fe} formalisms, have proven to be a powerful tool in the understanding of the theory at the quantum level. This stands in contrast to the more conventional Green-Schwarz (GS) and Ramond-Neveu-Schwarz (RNS) superstring descriptions \cite{Green:1987sp}. Even though the GS superstring has manifest spacetime SUSY, quantization becomes difficult due to the lack of manifest Lorentz-covariance in the light-cone gauge, and computations of scattering amplitudes from the GS superstring remain a challenging task. Despite the fact that the RNS superstring is quantizable in a Lorentz-covariant manner, spacetime supersymmetry is not manifest, and the theory has an infinite number of SUSY charges related by picture-changing \cite{Friedan:1985ge}.

For the superstring in a flat ten-dimensional background, the spacetime supersymmetric pure spinor formalism has been useful for constructing massive vertex operators \cite{Berkovits:2002qx} and in the computation of scattering amplitudes at the tree- and loop-level \cite{Gomez:2013sla} \cite{Berkovits:2005bt} \cite{Berkovits:2022ivl} \cite{Berkovits:2022fth} \cite{Mafra:2022wml}. In a curved ten-dimensional background, manifest spacetime supersymmetry of the formalism has enabled the construction of quantizable sigma-model actions in the presence of Ramond-Ramond states \cite{Berkovits:2001ue}, including the $\rm AdS_5 \times S^5$ background \cite{Berkovits:2004xu}.

When it comes to compactifications of the superstring on Calabi-Yau backgrounds, the so-called hybrid formalism for the superstring \cite{Berkovits:1996bf} plays a distinguished role. One of its appealing features is that quantization can be achieved while preserving some of the spacetime SUSYs in a manifest way. As opposed to the GS-action, the hybrid action is quadratic in a flat background. Additionally, the hybrid description enjoys an $\mathcal{N}=4$ superconformal symmetry, which can be used to compute $n$-point multiloop superstring amplitudes from a topological prescription \cite{Berkovits:1994vy}.

The hybrid description of the superstring consists of a field redefinition from the gauge-fixed RNS superstring into a set of GS-like variables, allowing spacetime supersymmetry to be made manifest. This can be achieved in either two dimensions \cite{Berkovits:2001tg}, four dimensions \cite{Berkovits:1996bf} \cite{Benakli:2021jxs}, six dimensions \cite{Berkovits:1994vy}, or in a $\rm U(5)$ subgroup of the ten-dimensional super-Poincaré group \cite{Berkovits:1999in}. Here, our focus will be the construction down to six-dimensional spacetime.

Since the six-dimensional hybrid formalism has four of the eight $\ta$ coordinates of superspace as fundamental worldsheet variables \cite{Howe:1983fr}, only half of the eight $d=6$ $\mathcal{N}=1$ SUSYs are manifest, i.e., act geometrically in the target-superspace. To overcome this issue, ref.~\cite{Berkovits:1999du} introduced four more $\ta$ coordinates, along with their conjugate momenta, as fundamental worldsheet fields together with four fermionic first-class constraints $D_{\al}$. In such a manner that the gauge symmetry generated by these constraints can be used to gauge away the new variables. Therefore, when $D_{\al}=0$, one recovers the hybrid description.

Under these circumstances, the constraint $D_{\al}=0$ has to be imposed ``by hand'', which means that identifying the usual $d=6$ $\mathcal{N}=1$ superfields in the vertex operator is not feasible in practice. Consequently, it is unclear where each of the component fields sits in the vertex before using $D_{\al}=0$ and making contact with the usual six-dimensional hybrid description. In addition, this also implies a major obstacle for defining scattering amplitudes with vertex operators depending on eight $\ta$s. The goal of this paper is to overcome some of these issues.

More specifically, in this work our objective is to formulate a quantizable formalism for the superstring with manifest $d=6$ $\mathcal{N}=1$ supersymmetry. The approach will feature a manifestly spacetime supersymmetric BRST operator, vertex operators, and a prescription for computing scattering amplitudes. In particular, the formalism incorporates the eight $\theta$ coordinates of $d=6$ $\mathcal{N}=1$ superspace as fundamental worldsheet variables, and unconstrained bosonic ghost fields $\lambda^{\alpha}$. The construction of a supersymmetric BRST charge, vertex operators, and a prescription for amplitudes together comprise the foundation for any complete and fully functional covariant description of the superstring.


The original results presented in this work include: (i) the calculation of the $\mathcal{N}=2$ supercurrent $G^+_{\rm hyb}$ in the six-dimensional hybrid formalism, with careful consideration of normal-ordering contributions; (ii) the construction of a manifestly supersymmetric compactification-independent vertex operator for the superstring that implies the $d=6$ $\mathcal{N}=1$ super-Yang-Mills (SYM) equations of motion in superspace; and (iii) the formulation of a manifestly $d=6$ $\mathcal{N}=1$ supersymmetric prescription for $n$-point scattering amplitudes.

The reasoning supporting the framework proposed in this manuscript is as follows. We will show that, after relaxing the harmonic constraint $D_{\al}$ of ref.~\cite{Berkovits:1999du}, ghost number one supersymmetric unintegrated vertex operators $U$ can be constructed in terms of $d=6$ $\mathcal{N}=1$ superfields. In addition, it will be proven that BRST invariance of $U$ implies the $d=6$ super-Yang-Mills equations of motion in superspace \cite{Howe:1983fr} \cite{Koller:1982cs}. Besides the fermionic fields $\ta$, unconstrained bosonic ghost-fields $\la^{\al}$, and its conjugate momenta, will be added to the worldsheet action in such a way that the total central charge of the stress-tensor $T$ vanishes. 

The BRST current of the theory will take the form $G^+ = G^+_{\rm hyb} - \la^{\al} D_{\al}$, where $G^+_{\rm hyb}$ is the positively charged $\mathcal{N}=2$ supercurrent of the hybrid formalism in supersymmetric notation, and the term $- \la^{\al} D_{\al}$ is responsible for the relaxation of the constraint $D_{\al}=0$. The ghost-number current will be defined in terms of the $\rm U(1)$ current of the hybrid $\mathcal{N}=2$ algebra. Furthermore, as in the non-minimal pure spinor formalism \cite{Berkovits:2005bt}, non-minimal/topological variables will be introduced to the BRST current in order to define supersymmetric scattering amplitude computations with a suitable regulator $\mathcal{R}$. We will end up by using the amplitude prescription to compute a three-point amplitude of $d=6$ SYM states.

The structure of the paper is as follows. In Section \ref{sec1}, we review the six-dimensional hybrid formalism in a flat background together with its description in terms of the harmonic constraint $D_{\al}$. The worldsheet variables, superconformal generators, and vertex operators of the theory are described in detail. In Section \ref{exthyb}, we define a new BRST operator $\oint G^+$ with the addition of the non-minimal variables, and construct a $d=6$ $\mathcal{N}=1$ supersymmetric unintegrated vertex operator $U$, and an integrated vertex operator $W$. It is then shown that BRST invariance of $U$ implies the $d=6$ SYM equations of motion in superspace. With both integrated and unintegrated vertex operators at our disposal, a tree-level scattering amplitude prescription is given which shares many similarities to the $d=10$ non-minimal pure spinor formalism one \cite{Berkovits:2005bt}. Section \ref{concl} contains our conclusion. There are a number of appendices where further technical details are given.

\section{Hybrid formalism in a flat six-dimensional background} \label{sec1}

In this section, we review the worldsheet variables and the physical state conditions of the hybrid formalism for the superstring in a flat six-dimensional background. Novel results include identity \eqref{Q4Gplus} and the computation of eq.~\eqref{gplusid} taking care of the normal-ordering contributions. This description will serve as the starting point for Section \ref{exthyb} of the paper. 

\subsection{Worldsheet action and superconformal generators} \label{half6dhybrid}

After performing a field redefinition of the gauge-fixed RNS variables \cite{Berkovits:1994vy} \cite{Berkovits:1999im}, the worldsheet fields of the six-dimensional part consist of six conformal weight zero bosons $x^{\ab}$, $\ab = \{0$ to $5\}$, and a canonically conjugate left-moving pair of fermions $\{ p_{\al }, \ta^{\al } \}$ of conformal weight one and zero, respectively, together with its right-moving part $\{ \widehat{p}_{\alh }, \tah^{\alh } \}$, where $\al , \alh =\{1$ to $4\}$. 

In a flat six-dimensional background the worldsheet action in conformal gauge takes the form\footnote{The OPEs between our fundamental worldsheet fields $\{p_{\al}, \ta^{\al}, \rho ,\sg\}$ are given by eqs.~\eqref{freefieldOPEs}, with the replacement of $\al j \rightarrow \al$.}
\begin{equation}
S=\int d^2z \, \bigg( \frac{1}{2} \pd x^{\ab} \pb x_{\ab} + p_{\al } \pb \ta^{\al } + \widehat{p}_{\alh } \pd \widehat{\ta}^{\alh } \bigg) + S_{\rho, \sg} + S_{C}\,, \label{wsactionflat0}
\end{equation}
where $\frac{\pd}{\pd z} = \pd$, $\frac{\pd}{\pd \zb}=\pb$, $S_{\rho, \sg}$ is the part of the action characterizing the chiral bosons $\rho$ and $\sigma$, as well as their anti-chiral counterparts, to be defined by their OPEs and stress-tensor below, and $S_{C}$ corresponds to the four-dimensional compactification variables. These variables can be taken to be any $c=6$ $\mathcal{N}=2$ superconformal field theory describing the compactification manifold, which can be either $\rm K3$ or $\rm T^4$  \cite{Berkovits:1999im}.

For the Type-IIB (Type-IIA) superstring, an up $\al$ index and an up (down) $\alh$ index transform as a Weyl spinor of SU(4), a down $\al$ index and a down (up) $\alh$ index transform as an anti-Weyl spinor of SU(4). In this case, note that Weyl and anti-Weyl spinors are not related by complex conjugation. Also, we will only discuss the open string part of the worldsheet theory in what follows.

To define physical states, one needs to supplement the action \eqref{wsactionflat0} with the twisted $c=6$ $\mathcal{N}=2$ constraints \cite{Berkovits:1994vy}
\begin{subequations} \label{BVWconstraints}
\begin{align}
T_{\rm hyb}&= -  \frac{1}{2} \pd x^{\ab} \pd x_{\ab} - p_{\al} \pd \ta^{\al}  - \frac{1}{2} \pd \rho \pd \rho - \frac{1}{2} \pd \sg \pd \sg  + \frac{3}{2}\pd^2 ( \rho + i \sg) + T_C\,,  \\
G^+_{\rm hyb} & =  -(p)^4 e^{-2\rho -i \sg} + \frac{i}{2}p_{\al}p_{\bt} \pd x^{\al \bt} e^{-\rho}  -\frac{1}{2} \pd x^{\ab} \pd x_{\ab}e^{i \sg}  - p_{\al} \pd \ta^{\al}e^{i \sg}  \label{Gplushyb} \nn \\
& - \frac{1}{2}\pd ( \rho + i \sg) \pd ( \rho + i \sg) e^{i\sg} + \frac{1}{2}\pd^2 ( \rho + i \sg) e^{i \sg} + G^+_C \,, \\
G^{-}_{\rm hyb} & = e^{-i \sg} + G^{-}_{C}\,, \\
J_{\rm hyb} &= \pd ( \rho + i \sg) + J_C\,,
\end{align}
\end{subequations}
where $(p)^4 = \frac{1}{24} \ep^{\al \bt \g \dt} p_{\al} p_{\bt} p_{\g} p_{\dt}$, $x_{\al \bt} = \sigma^{\ab}_{\al \bt} x_{\ab}$ and $\sg^{\ab}_{\al \bt}$ are the six-dimensional Pauli matrices, which are $4 \times 4$ antisymmetric in the spinor indices. In this work, we use the same conventions for the six-dimensional Pauli matrices as in ref.~\cite[Appendix A]{Daniel:2024kkp}. 

Note that $\{T_C,G^\pm_C,  J_C \}$ represent a twisted $c=6$ $\mathcal{N}=2$ superconformal field theory describing the compactification manifold, so that $\{T_{\rm hyb}-T_C,G^{\pm}_{\rm hyb} -G^\pm_C,  J_{\rm hyb} -J_C \}$ describe a $c=0$ $\mathcal{N}=2$ superconformal algebra (SCA). The generators $\{T_C,G^\pm_C,  J_C \}$ have no poles with the six-dimensional worldsheet variables and no poles with the chiral bosons $\{ \rho, \sg \}$. For the closed string, we also have the right-moving piece of the above algebra. 

The operators $e^{m\rho + n i \sg}$ are conformal tensors and have conformal weight $\frac{1}{2}( - m^2 + 3m + n^2 - 3n) $. The definition of normal-ordering used in eqs.~\eqref{BVWconstraints}, and in the rest of this work, is presented in Appendix \ref{normalO}. In particular, notice that we can write the first two terms in the second line of \eqref{Gplushyb} in a more compact form as 
\begin{align}
- \frac{1}{2}\pd ( \rho + i \sg) \pd ( \rho + i \sg) e^{i\sg} + \frac{1}{2}\pd^2 ( \rho + i \sg) e^{i \sg} & = \big( \pd e^{-\rho -i \sg}, e^{\rho + 2i \sg} \big) \,,
\end{align}
using our normal-ordering prescription.

Correspondingly, any twisted $c=6$ $\mathcal{N}=2$ SCA \eqref{BVWconstraints} can be extended to a twisted small $c=6$ $\mathcal{N}=4$ SCA \cite{Berkovits:1994vy} by adding two bosonic currents and two supercurrents, as detailed in Appendix \ref{SCAs}. The additional $\mathcal{N}=4$ superconformal generators in the six-dimensional hybrid formalism are
\begin{subequations}
\begin{align}
\widetilde{G}^+_{\rm hyb} & = e^{\rho} J^{++}_C - e^{\rho + i \sg} \widetilde{G}^+_C \label{BVgtildeplus} \,, \\
\widetilde{G}^-_{\rm hyb} & =\bigg( -(p)^4 e^{-3 \rho -2 i \sg} + \frac{i}{2}p_{\al}p_{\bt} \pd x^{\al \bt}e^{-2 \rho -i \sg}   -\frac{1}{2} \pd x^{\ab} \pd x_{\ab}e^{-\rho} - p_{\al} \pd \ta^{\al}  e^{-\rho} \nn \\
&- \big(\pd e^{-\rho -i \sg}, e^{i\sg}\big) \bigg) J^{--}_C + e^{-\rho -i \sg} \widetilde{G}^-_C \,, \\
J^{++}_{\rm hyb} & = - e^{\rho + i \sg} J^{++}_C \,, \\
J^{--}_{\rm hyb} & = e^{-\rho -i \sg} J^{--}_C\,,
\end{align}
\end{subequations}
where $\{\widetilde{G}^\pm_C,J^{\pm \pm}_C\}$, that together with $\{T_C,G^\pm_C,  J_C\}$, form a twisted small $c=6$ $\mathcal{N}=4$ SCA which has no poles with the six-dimensional worldsheet variables and also no poles with the chiral bosons $\{\rho, \sg \}$.

The spacetime supersymmetry charges in the six-dimensional hybrid formalism are given by \cite{Berkovits:1999im}
\begin{align}\label{SUSYshyb}
Q^{\rm hyb}_{\al 1} & = \oint p_{\al} \,,& Q^{\rm hyb}_{\al 2} & = \oint \big( e^{-\rho -i \sg} p_{\al} + i \pd x_{\al \bt} \ta^{\bt} \big) \,,
\end{align}
and satisfy the spacetime SUSY algebra $\{ Q^{\rm hyb}_{\al 1}, Q^{\rm hyb}_{\al 2} \} = -i \oint \pd x_{\al \bt}$. Note that the charge $Q^{\rm hyb}_{\al 2}$ has the presence of the $\{\rho, \sg\}$-ghosts and, for that reason, it is called the ``non-standard'' supersymmetry generator \cite{Berkovits:1999im}.

The superconformal generators \eqref{BVWconstraints} are manifestly invariant under the SUSY charge $Q^{\rm hyb}_{\al 1}$. Invariance under $Q^{\rm hyb}_{\al 2}$ is difficult to check for the supercurrent $G^+_{\rm hyb}$. However, the latter can be made manifest by noting that one can write the supercurrent as
\begin{align}\label{Q4Gplus}
G^+_{\rm hyb} & = - \frac{1}{24} \ep^{\al \bt \g \dt} Q^{\rm hyb}_{\al 2} Q^{\rm hyb}_{\bt 2} Q^{\rm hyb}_{\g 2} Q^{\rm hyb}_{\dt 2} e^{2 \rho + 3 i \sg}  + G^+_C\,,
\end{align}
which is a property that also holds in an $\rm AdS_3 \times S^3$ background including the normal-ordering contributions \cite{Daniel:2024kkp}.

Note that we are denoting operators defined throughout this section with the subscript/superscript ``$\rm hyb$'', so as to not cause confusion with the generators to be introduced in Section \ref{exthyb}.

\subsection{Physical states}

Following refs.~\cite{Berkovits:1999im} \cite{Daniel:2024kkp}, physical states $V_{\rm hyb}$ of the theory are defined to satisfy the equation of motion\footnote{For a holomorphic operator $\mathcal{O}$ with conformal dimension $h$,  $(\mathcal{O})_r$ is defined by the usual mode expansion in the plane, namely, $\mathcal{O}(z) = \sum_r (\mathcal{O})_rz^{-r - h}$.}
\begin{align}
(G^+_{\rm hyb})_0  (\widetilde{G}^+_{\rm hyb})_0 V_{\rm hyb} & = 0  \,, 
\end{align}
so that the vertex operator $V_{\rm hyb}$ is defined up to the gauge transformation
\begin{align}\label{gaugetransfVhyb}
\dt V_{\rm hyb} & = (G^+_{\rm hyb})_0 \La + (\widetilde{G}_{\rm hyb})_0 \Omega\,,
\end{align}
for some $\La$ and some $\Omega$. Moreover, it is consistent to impose the additional gauge-fixing conditions
\begin{align}
(G^-_{\rm hyb})_0 V_{\rm hyb} & = (\widetilde{G}^-_{\rm hyb})_0 V_{\rm hyb} = (T_{\rm hyb})_0 V_{\rm hyb} = (J_{\rm hyb})_0 V_{\rm hyb} = 0\,.
\end{align}

As an example, let us consider the massless compactification-independent states in six dimensions for the open superstring. The most general vertex operator with conformal weight zero and no poles with the $\rm U(1)$-current $J_{\rm hyb}$ has the form
\begin{align}\label{Vhyb}
V_{\rm hyb} & = \sum_{n=0}^{\infty} V_n e^{n (\rho + i \sg)} \,.
\end{align}

The conditions of no double poles or higher with $G^-_{\rm hyb}$ and with $\widetilde{G}^-_{\rm hyb}$ imply that $V_n =0$ for $n \geq 2$ and $n \leq -2$, respectively. From the remaining equations coming from $ (\widetilde{G}^-_{\rm hyb})_0 V_{\rm hyb} = 0$, together with the gauge transformations \eqref{gaugetransfVhyb}, one can gauge-fix $V_{\rm hyb}$ to the form
\begin{align}\label{Vhybghoststruc}
V_{\rm hyb} & = V_1 e^{\rho + i \sg} + V_0 \,,
\end{align}
where 
\begin{subequations}
\begin{align}
V_1 & = \ta^{\al} \chi_{\al 2} + \frac{i}{2} (\ta \sg^{\ab} \ta) a_{\ab} -(\ta^3)_{\al} \psi^{\al 2}\,, \label{superfieldV1}\\
V_0 & = \ta^{\al} \chi_{\al 1} \,,
\end{align}
\end{subequations}
with $\psi^{\al j} = \ep^{jk} i \pd^{\al \bt} \chi_{\bt k}$ the gluino and $a_{\ab}$ the gluon. The two-dimensional Levi-Civita symbol takes the values $\ep_{12} = \ep^{21} = 1$. Even though $\chi_{\al j}$ is not gauge-invariant, we have that $\dt \psi^{\al j} =0$ under a gauge transformation. In our conventions, we are using
\begin{align}
(\ta^3)_{\al} & = \frac{1}{6} \ep_{\al \bt \g \dt} \ta^{\bt} \ta^{\g} \ta^{\dt} \,,& (\ta^4) & = \frac{1}{24} \ep_{\al \bt \g \dt} \ta^{\al} \ta^{\bt} \ta^{\g} \ta^{\dt} \,,
\end{align}
where $\ep_{\al \bt \g \dt}$ is the Levi-Civita symbol with $\ep_{1234}=1$.

The superfield $V_1$ satisfies the equation of motion $\pd^{\al \bt} \nabla_{\al} \nabla_{\bt} V_1 =0$ which, together with the gauge transformations, imply that the component fields obey
\begin{subequations}\label{fieldcontentSYM}
\begin{align}
\pd^{\ab}\pd_{\ab} a_{\bb}  = \pd^{\ab} a_{\ab} & = 0 \,,& \dt a_{\ab} &= \pd_{\ab} \la \,, \\
\pd_{\al \bt} \psi^{\bt j} & = 0 \,, &  \dt \psi^{\al j} & = 0 \,,
\end{align}
\end{subequations}
for some $\la$. The gauge transformation of $a_{\ab}$ comes from choosing $\La = (\ta)^4 \la$ in \eqref{gaugetransfVhyb}.

Eqs.~\eqref{fieldcontentSYM} are the field content of $d=6$ super-Yang-Mills (SYM). It is also important to note that all degrees of freedom are contained in the superfield $V_1$ of eq.~\eqref{superfieldV1}. In Section \ref{exthyb}, we will see how one can describe superstring vertex operators for the SYM states in terms of the usual superfields of $d=6$ $\mathcal{N}=1$ superspace \cite{Howe:1983fr}.

Furthermore, by considering \eqref{Vhyb} in the gauge where $(G_{\rm hyb}^+)_0 (\widetilde{G}_{\rm hyb}^+)_0 V_{\rm hyb} =0$, we find that the integrated vertex operator for the open superstring compactification-independent massless states is 
\begin{align}
W_{\rm hyb} & = \int (G^+_{\rm hyb})_0 (G^-_{\rm hyb})_{-1} V_{\rm hyb} \nn \\
& = \int \bigg( - e^{- \rho -i \sg} p_{\al} (\nabla^3)^{\al} - \frac{i}{2} \pd x^{\al \bt} \nabla_{\al } \nabla_{\bt} + i p_{\al} \pd^{\al \bt} \nabla_{\bt} \bigg) V_1 + p_{\al} (\nabla^3)^{\al} V_2\,.
\end{align}

\subsection{Six-dimensional hybrid formalism with harmonic-like constraints} \label{secharmonic}

Even though the six-dimensional hybrid formalism presented above preserves manifest $\rm SO(1,5)$ Lorentz invariance, only half of the eight supersymmetries of $d=6$ $\mathcal{N}=1$ superspace are manifest, i.e., act geometrically in the target superspace. This can be observed by the fact that only four left-moving $\ta$s are present in the worldsheet action \eqref{wsactionflat0} as fundamental fields. 

However, one can proceed as in ref.~\cite{Berkovits:1999du} and add four more left-moving $\ta$s and four right-moving $\tah$s, as well as their conjugate momenta as fundamental worldsheet variables to the action. This doubling of fermionic degrees of freedom can be accomplished by appending the index $j=\{1,2\}$ to $\{p_{\al}, \ta^{\al}\}$, so that we end up with
\begin{equation}
S=\int d^2z \, \bigg( \frac{1}{2} \pd x^{\ab} \pb x_{\ab} + p_{\al j} \pb \ta^{\al j} + \widehat{p}_{\alh j} \pd \widehat{\ta}^{\alh j} \bigg) + S_{\rho, \sg} + S_{C}\,, \label{wsactionflat}
\end{equation}
where $p_{\al j}= \ep_{jk} p^k_{\al}$, $\ep_{12}=-\ep^{12}=1$, $\ep_{jk} \ep^{kl}=\dt_j^l$ and repeated indices are summed over. 

Consequently, eq.~\eqref{wsactionflat} is invariant under the $d=6$ $\mathcal{N}=1$ spacetime supersymmetry transformations generated by the charge
\begin{align} \label{susygen1}
Q_{\al j} & = \oint \bigg( p_{\al j} - \frac{i}{2} \ep_{jk} \pd x_{\al \bt} \ta^{\bt k} - \frac{1}{24} \ep_{\al \bt \g \dt} \ep_{jk} \ep_{lm}\ta^{\bt k} \ta^{\g l} \pd \ta^{\dt m} \bigg) \,,
\end{align}
and which satisfy the $d=6$ $\mathcal{N}=1$ SUSY algebra
\begin{align}
\{ Q_{\al j}, Q_{\bt k} \} & = -i \ep_{jk} \oint \pd x_{\al \bt}\,.
\end{align}
For the closed string, we also have a left- and a right-moving supersymmetry generator $Q_{\al j}$ and $\widehat{Q}_{\alh j}$, respectively. These charges then generate the $d=6$ $\mathcal{N}=2$ supersymmetry and, hence, for Type II strings the amount of SUSY is doubled. 

Beyond that, it is convenient to construct extensions of the worldsheet fields $\{p_{\al j}, \pd x^{\ab} \}$ that are invariant under the transformations generated by \eqref{susygen1}. One can easily check that this is achieved by the following on-shell spacetime supersymmetric --- or just supersymmetric --- worldsheet variables
\begin{subequations}\label{susyvars}
\begin{align}
d_{\al j}&=p_{\al j} + \frac{i}{2}\ep_{jk} \pd x_{\al \bt} \ta^{\bt k} + \frac{1}{8} \ep_{\al \bt \g \dt} \ep_{jk} \ep_{lm} \ta^{\bt k} \ta^{\g l} \pd \ta^{\dt m}\,, \\
\Pi^{\ab} &= \pd x^{\ab} - \frac{i}{2}\ep_{jk} \sg^{\ab}_{\al \bt} \ta^{\al j} \pd \ta^{\bt k}\,.
\end{align}
\end{subequations}

The six-dimensional worldsheet fields in \eqref{wsactionflat} have the following singularities in their OPEs
\begin{subequations}\label{freefieldOPEs}
\begin{align}
p_{\al j}(y)\ta^{\bt k}(z) & \sim \dt^k_j \dt^\bt_\al (y-z)^{-1} \,, \\
\pd x^{\ab}(y) \pd x^{\bb}(z) & \sim - \eta^{\ab \bb}(y-z)^{-2}\,, \\
\rho(y)\rho(z) &\sim - \log(y-z)\,, \\
\sg(y)\sg(z) &\sim - \log(y-z)\,,
\end{align}
\end{subequations}
where $\eta^{\ab \bb}= \text{diag}(-,+,+,+,+,+)$ and, in turn, eqs.~\eqref{freefieldOPEs} can be used to show that the supersymmetric variables \eqref{susyvars} satisfy
\begin{subequations} \label{susyOPEs}
\begin{align}
d_{\al j}(y) d_{\bt k}(z) &\sim (y-z)^{-1} i \ep_{jk} \Pi_{\al \bt}(z)\,, \\
d_{\al j}(y) \Pi^{\ab}(z) &\sim -(y-z)^{-1} i\ep_{jk} \sg^{\ab}_{\al \bt} \pd \ta^{\bt k}(z)\,,\\
\Pi^{\ab}(y)\Pi^{\bb}(z) &\sim - (y-z)^{-2} \eta^{\ab \bb}\,, \\
d_{\al j}(y) \pd \ta^{\bt k}(z) &\sim (y-z)^{-2} \dt^k_j \dt^\bt_\al\,.
\end{align}
\end{subequations}
Also, notice the following ordering effect using the OPEs of the fundamental fields
\begin{subequations}\label{dPi}
\begin{align}
\oint \frac{dy}{y-z} \, \Pi_{\al \bt}(y) d_{\g j}(z) &=  \frac{3i}{4} \ep_{jk} \ep_{\al \bt \g \dt} \pd^2 \ta^{\dt k}(z)\,, \\
\oint \frac{dy}{y-z} \, d_{\g j}(y)\Pi_{\al \bt}(z) &= -\frac{i}{4} \ep_{jk} \ep_{\al \bt \g \dt} \pd^2 \ta^{\dt k}(z)\,.
\end{align}
\end{subequations}

As we have argued, the worldsheet action \eqref{wsactionflat} is invariant under the $d=6$ $\mathcal{N}=1$ spacetime supersymmetry transformations, however, in order to preserve the description of the original six-dimensional hybrid superstring, one must include a set of constraints which reduce the action \eqref{wsactionflat} to \eqref{wsactionflat0}. This can be accomplished by the fermionic first-class constraints \cite{Berkovits:1999du}
\begin{align}
D_{\al} & = d_{\al 2} - e^{-\rho -i \sg} d_{\al 1}\,, \label{harmonic}
\end{align}
and since
\begin{align}\label{commDs}
[ D_{\al}, \ta^{\bt 2} ] & = \dt^{\bt}_{\al}\,,
\end{align}
one can use \eqref{harmonic} to gauge-fix \eqref{wsactionflat} to \eqref{wsactionflat0}. Therefore, working with the action \eqref{wsactionflat} and the harmonic constraint $D_{\al}$, it is possible to manifestly preserve all of the $d=6$ $\mathcal{N}=1$ supersymmetries.\footnote{We note in passing that there exists a similarity transformation with the property $e^S d_{\al 2}e^{-S} = D_{\al}$ where $S= \ta^{\al 2} d_{\al 1}e^{-\rho -i\sg} - \frac{i}{2} \ta^{\al 2} \ta^{\bt 2} \Pi_{\al \bt} e^{-\rho -i\sg} + (\ta^2)^3_{\al} \pd \ta^{\al 1}e^{-\rho -i\sg} + \frac{1}{2} (\ta^2)^4 \pd (\rho + i \sg) e^{-2\rho -2i\sg}$.}

In this case, the $\mathcal{N}=2$ constraints \eqref{BVWconstraints} are modified and can be written in a manifestly spacetime supersymmetric form as \cite{Berkovits:1999du}
\begin{subequations}\label{SCAgen}
\begin{align}
T_{\rm hyb}&= -  \frac{1}{2} \Pi^{\ab} \Pi_{\ab} - d_{\al 1} \pd \ta^{\al 1} - e^{-\rho -i \sg} d_{\al 1} \pd \ta^{\al 2}  - \frac{1}{2} \pd \rho \pd \rho - \frac{1}{2} \pd \sg \pd \sg  \nn \\
& + \frac{3}{2}\pd^2 ( \rho + i \sg) + T_C\,, \label{stressT} \\
G^+_{\rm hyb} & = - (d_1)^4 e^{-2\rho -i \sg} + \frac{i}{2}d_{\al 1}d_{\bt 1} \Pi^{\al \bt}e^{-\rho} + d_{\al1} \pd \ta^{\al 2}\pd (\rho + i \sg)e^{-\rho} +  d_{\al 1} \pd^2 \ta^{\al 2}e^{-\rho}  \nn \\
&  -\frac{1}{2}\Pi^{\ab} \Pi_{\ab} e^{i \sg} - d_{\al 1}\pd \ta^{\al 1} e^{i \sg} - \frac{1}{2}\pd ( \rho + i \sg) \pd ( \rho + i \sg) e^{i\sg}  \nn \\
&  + \frac{1}{2}\pd^2 ( \rho + i \sg) e^{i \sg} + G^+_C \,, \\
G^{-}_{\rm hyb} & = e^{-i \sg} + G^{-}_{C}\,, \\
J_{\rm hyb} &= \pd ( \rho + i \sg) + J_C\,, \label{Jhyb}
\end{align}
\end{subequations}
which, as in the previous section, still obey a twisted $c=6$ $\mathcal{N}=2$ SCA, and we defined $(d_1)^4 = \frac{1}{24}\ep^{\al \bt \g \dt}d_{\al 1} d_{\bt 1} d_{\g 1} d_{\dt 1}$. Let us mention that when one gauge fix $\ta^{\al 2}  =0$, the constraints \eqref{SCAgen} reduce to the ones compatible with the action \eqref{wsactionflat0}, i.e., eqs.~\eqref{BVWconstraints}. Note that the stress-tensor $T_{\rm hyb}$ is the expected stress tensor when $D_{\al}=0$, because 
\begin{equation}\label{nonstress-tensor}
 - d_{\al 1} \pd \ta^{\al 1} - e^{-\rho -i \sg} d_{\al 1} \pd \ta^{\al 2} = - d_{\al j} \pd \ta^{\al j} +D_{\al} \pd \ta^{\al 2}\,.
\end{equation}

It is also important to be aware that the $\mathcal{N}=2$ algebra is preserved independently of how one chooses to gauge-fix the local symmetry generated by $D_{\al}$. This is because the form of the $\mathcal{N}=2$ generators \eqref{SCAgen} was chosen so that they have no poles with the harmonic-like constraint \eqref{harmonic}. The non-trivial part in showing this is for the generator $G^+_{\rm hyb}$, nonetheless, it becomes manifest by noting the property that one can write $G^+_{\rm hyb}$ as
\begin{equation}
G^+_{\rm hyb} =-\frac{1}{24} \ep^{\al \bt \g \dt} [D_{\al},\{ D_{\bt},[ D_{\g},\{ D_{\dt},e^{2 \rho +3 i\sg}\}]\}] + G_C^+\,, \label{gplusid}
\end{equation}
where the graded bracket $[D_{\al}, \mathcal{O} \}(z) = \oint dy\, D_{\al}(y) \mathcal{O}(z)$ denotes the simple pole in the OPE between $D_{\al}$ and $\mathcal{O}$. 

The details of the calculation establishing eq.~\eqref{gplusid} are given in Appendix \ref{gpluscalc}. To the knowledge of the author, this is the first time that eq.~\eqref{gplusid} is proven considering the normal-ordering contributions. Note also the similarity between identities \eqref{gplusid} and \eqref{Q4Gplus}.

For the massless compactification-independent sector of the open superstring, the vertex operator now reads
\begin{align}\label{BVvertex2}
V_{\rm hyb}= \sum_{n = -\infty}^{\infty}  V_n(x, \theta)e^{n(\rho + i \sg)}\,,
\end{align}
which takes the same form as in eq.~\eqref{Vhyb}, but now $V_n(x,\ta)$ depends on the zero modes of $\{x^{\ab}, \ta^{\al j}\}$. Therefore, it contains the eight fermionic $\ta$ coordinates of $d=6$ $\mathcal{N}=1$ superspace. Of course, contrasting with the hybrid formalism of the previous section, in the present case the physical states $V_{\rm hyb}$ also have to be annihilated by $D_{\al}$.

It is interesting to note what is the effect of imposing the constraint $D_{\al}$ for $V_{\rm hyb}$ of eq.~\eqref{BVvertex2}. To do that, let us first define the new superspace variables
\begin{align}
\ta^{\al -} & = \frac{1}{2} \big( \ta^{\al 2} - e^{\rho + i \sg} \ta^{\al 1} \big)\,, & \ta^{\al +} & = \frac{1}{2}\big( \ta^{\al 1} + e^{-\rho - i \sg} \ta^{\al 2} \big) \,.
\end{align}
The condition that $V_{\rm hyb}$ has no poles with $D_{\al}$ implies that
\begin{align}
\big( \nabla_{\al 2} - e^{-\rho -i \sg} \nabla_{\al 1} \big) V_{\rm hyb} & = 0\,, \label{constraintVzeromode}
\end{align}
where $\nabla_{\al j} = \frac{\pd}{\pd \ta^{\al j}} - \frac{i}{2} \ep_{jk} \ta^{\bt k}\pd_{\al \bt} $ is the zero mode of $d_{\al j}$ acting on $V_{\rm hyb}$. By defining ${x^\prime}^{\ab} = x^{\ab}$ and then doing the shift
\begin{align}
{x^\prime}^{\ab} + i \ta^{\al -} \ta^{\bt +} \sg^{\ab}_{\al \bt} \mapsto x^{\ab} \,,
\end{align}
we learn that $V_{\rm hyb}$ is independent of $\ta^{\al -}$ and, for that reason, it is a function of only the zero modes of $\{x^{\ab},\ta^{\al +} \}$. As a consequence, after identifying $\ta^{\al +} = \ta^{\al}$ the component fields of $V_{\rm hyb}$ in \eqref{BVvertex2} can be related to the component fields of $V_{\rm hyb}$ in \eqref{Vhyb}, therefore, we recover the usual six-dimensional description of the vertex operator in Section \ref{half6dhybrid}. Nonetheless, the identification of the component fields is only possible after imposing $D_{\al}=0$.

From eqs.~\eqref{SCAgen}, one can construct the remaining twisted small $c=6$ $\mathcal{N}=4$ generators, and in the gauge where $(G_{\rm hyb}^+)_0 (\widetilde{G}_{\rm hyb}^+)_0 V_{\rm hyb} =0$ the integrated vertex operator for the massless compactification-independent states of the open superstring now reads \cite{Berkovits:1999du}
\begin{align}\label{6dintvertexsec1}
W_{\rm hyb} & = \int (G^+_{\rm hyb})_0 (G^-_{\rm hyb})_{-1} V_{\rm hyb} \nn \\
& = \int \, \bigg[ \bigg( - \frac{1}{6}e^{-\rho -i \sg} \ep^{\al \bt \g \dt} d_{\al 1} \nabla_{\bt 1} \nabla_{\g 1} \nabla_{\dt 1} - \frac{i}{2} \Pi^{\al \bt} \nabla_{\al 1} \nabla_{\bt 1} \nn \\
& + i d_{\al 1} \pd^{\al \bt} \nabla_{\bt 1}  - \pd \ta^{\al 2} \nabla_{\al 1} \bigg) V_1+ \frac{1}{6}\ep^{\al \bt \g \dt} d_{\al 1} \nabla_{\bt 1} \nabla_{\g 1} \nabla_{\dt 1}V_2 \bigg] \nn \\
& = \int \,\bigg[ \frac{1}{6} \ep^{\al \bt \g \dt} \Big( d_{\al 2} \nabla_{\bt 1} \nabla_{\g 1} \nabla_{\dt 2} - d_{\al 1 }\nabla_{\bt 2} \nabla_{\g 2} \nabla_{\dt 1} \Big) \nn \\
&  + \frac{i}{4} \Pi^{\al \bt} [\nabla_{\al 1}, \nabla_{\bt 2}] -\frac{1}{2} \pd \ta^{\al 1} \nabla_{\al 1} + \frac{1}{2} \pd \ta^{\al 2} \nabla_{\al 2} \bigg]V_0\,,
\end{align}
where the supersymmetric derivative $\nabla_{\al j}$ satisfy the algebra $\{\nabla_{\al j} , \nabla_{\bt k} \}= -i \ep_{jk} \pd_{\al \bt}$. To arrive at the last equality in \eqref{6dintvertexsec1}, we subtracted a total derivative and used the relations implied by the constraint \eqref{harmonic}, namely, $\nabla_{\al 1} V_1 = - \nabla_{\al 2} V_0$ and $\nabla_{\al 1} V_2 = \nabla_{\al 2} V_1$.

\section{Extended hybrid formalism} \label{exthyb}

In spite of the fact that we have described the worldsheet action, $\mathcal{N}=2$ constraints and compactification-independent vertex operators while preserving manifest $d=6$ $\mathcal{N}=1$ supersymmetry in Section \ref{secharmonic}, it remains unclear what are the rules to compute correlation functions using the superconformal generators and vertex operators depending on all eight $\ta$ coordinates of $d=6$ $\mathcal{N}=1$ superspace. 

In addition, it is not evident if there is a relation between the vertex operator \eqref{BVvertex2} and the superfields appearing in superspace descriptions of $d=6$ SYM  \cite{Howe:1983fr} \cite{Koller:1982cs}. As a consequence, one cannot identify what each component of the superfield \eqref{BVvertex2} corresponds to before using the constraint $D_{\al}=0$ to make contact with \eqref{Vhyb}, which depends on only half of the $\ta$s. One of the purposes of this section is to clarify and understand how one can overcome these drawbacks by relaxing the constraint $D_{\al}=0$ in the definition of physical states.

\subsection{Worldsheet variables} \label{NMworldsheetvariables}
 
To the worldsheet theory \eqref{wsactionflat}, we introduce a bosonic spinor $\la^{\al}$ of conformal weight zero and its conjugate momenta $w_{\al}$ of conformal weight one. As we will momentarily see, the ghost $\la^{\al}$ will be responsible for relaxing the constraint $D_{\al}$. We also include the non-minimal variables $\{\lb_{\al}, r_{\al}\}$ \cite{Berkovits:2005bt} of conformal weight zero, as well as their conjugate momenta $\{ \wb^{\al}, s^{\al} \}$ of conformal weight one. The fields $\{s^{\al},r_{\al}\}$ are worldsheet fermions and $\{\wb^{\al}, \lb_{\al} \}$ bosons. 

The worldsheet action now takes the form
\begin{align}\label{exthybaction}
S & = \int d^2 z \, \bigg( \frac{1}{2} \pd x^{\ab} \pb x_{\ab} + p_{\al j} \pb \ta^{\al j} + w_{\al} \pb \la^{\al}+  s^{\al} \pb r_{\al} + \wb^{\al} \pb \lb_{\al} \nn \\
& + \widehat{p}_{\alh j} \pd \widehat{\ta}^{\alh j} + \wh_{\alh} \pd \lh^{\alh }  + \widehat{s}^{\alh} \pd \widehat{r}_{\alh} + \widehat{\wb}^{\alh} \pd \widehat{\lb}_{\alh} \bigg)  + S_{\rho, \sg} + S_C \,,
\end{align}
where the ``hatted'' fields are right-moving and, for simplicity, will be ignored in what follows. The singularities in the OPEs of the new variables are
\begin{subequations}
\begin{align}
w_{\al}(y) \la^{\bt}(z) & \sim - \dt^{\bt}_{\al}(y-z)^{-1} \,, \\
\wb^{\al}(y) \lb_{\bt}(z) & \sim - \dt^{\al}_{\bt}(y-z)^{-1} \,, \\
s^{\al}(y) r_{\bt}(z) & \sim  \dt^{\al}_{\bt}(y-z)^{-1} \,,
\end{align} 
 \end{subequations}
and, unlike in \cite{Berkovits:2005bt}, $\{\la^{\al}, \lb_{\al}, r_{\al}\}$ are not constrained. Note further that, as opposed to the worldsheet action \eqref{wsactionflat}, the stress-tensor of \eqref{exthybaction} has vanishing central charge.

\subsection{Extended twisted $c=6$ $\mathcal{N}=2$ generators}
With these additional variables, it is still possible to construct superconformal generators satisfying a twisted $c=6$ $\mathcal{N}=2$ SCA as in Section \ref{sec1}. 

In this case, we have
\begin{subequations}\label{NMSCA}
\begin{align}
T & = T_{\rm hyb} -D_{\al} \pd \ta^{\al 2} - w_{\al} \pd \la^{\al} - \wb^{\al} \pd \lb_{\al} - s^{\al} \pd r_{\al} \,, \label{NMstress-tensor} \\
G^+ & = G^+_{\rm hyb} - \la^{\al} D_{\al} - \wb^{\al} r_{\al} \,, \label{NMGplus}\\
G^- & = G^-_{\rm hyb} + w_{\al} \pd \ta^{\al 2} + s^{\al} \pd \lb_{\al} \,, \\
J & = J_{\rm hyb} - w_{\al} \la^{\al} - s^{\al} r_{\al}  \,, \label{NMJ}
\end{align}
\end{subequations}
where $\{T_{\rm hyb}, G^+_{\rm hyb}, G^-_{\rm hyb}, J_{\rm hyb} \}$ are the $c=6$ $\mathcal{N}=2$ generators of eqs.~\eqref{SCAgen}. Note that $T$ is now the usual stress-tensor, because the terms added in \eqref{NMstress-tensor} to $T_{\rm hyb}$ precisely cancel the atypical contribution in \eqref{nonstress-tensor}. Explicitly, we now have
\begin{align}
T&= -  \frac{1}{2} \Pi^{\ab} \Pi_{\ab} - d_{\al j} \pd \ta^{\al j}  - w_{\al} \pd \la^{\al} - \wb^{\al} \pd \lb_{\al} - s^{\al} \pd r_{\al}  \nn \\
& - \frac{1}{2} \pd \rho \pd \rho - \frac{1}{2} \pd \sg \pd \sg  + \frac{3}{2}\pd^2 ( \rho + i \sg) + T_C\,.
\end{align}

Of course, the superconformal generator $G^+$ continues to be nilpotent. This is easy to see from that fact that $G^+_{\rm hyb}$ has no poles with itself, no poles with $D_{\al}$ and the constraint $D_{\al}$ is first-class. 

It is important to comment on the significance of each of the contributions appearing in the fermionic generator $G^+$. The zero mode of $G^+_{\rm hyb}$ is related to the BRST operator $Q_{\rm RNS}$ of the RNS formalism in the gauge where $\ta^{\al 2} = 0$, this follows from the fact that the hybrid variables are related to the gauge-fixed RNS variables through a field redefinition \cite{Berkovits:1999im}. 

The term $- \la^{\al} D_{\al}$ in $G^+$ is necessary for the reason that we are relaxing the constraint $D_{\al}$. As a consequence, the condition $D_{\al} = 0$ from \eqref{harmonic} does not need to be imposed ``by hand'' in our definition of physical states from now on (see Section \ref{secNMvertex}). The last contribution, $-\wb^{\al} r_{\al}$, is the non-minimal/topological term \cite{Berkovits:2005bt}, and it implies that the cohomology of $(G^+)_0$ is independent of $\{\wb^{\al}, \lb_{\al}, s^{\al}, r_{\al} \}$ through the usual quartet argument. This term is required in order to get a $c=6$ $\mathcal{N}=2$ SCA  and it will play a key role in defining a spacetime supersymmetric prescription for scattering amplitude computations in Section \ref{scattamps}.

We should emphasize that even though we have an $\mathcal{N}=2$ SCA with critical central charge ($c=6$) in eqs.~\eqref{NMSCA}, the physical states of the superstring cannot be defined as $\mathcal{N}=2$ primaries like in the hybrid formalism \cite{Berkovits:1996bf}. The reason for this is because, by the quartet machanism, the cohomology of $(G^+)_0$ is guaranteed to be independent of the non-minimal/topological variables \cite{Berkovits:2005bt}. However, this mechanism has nothing to say about the primaries of the $\mathcal{N}=2$ algebra, i.e., if the they are preserved or not after the worldsheet theory is modified. Therefore, when studying vertex operators of the superstring, one must look for states in the cohomology of $(G^+)_0$.

As an additional observation, let us sketch a direct way to arrive at the supercurrent \eqref{NMGplus} from the six-dimensional hybrid formalism: by adding non-minimal variables and performing a suitable similarity transformation. Start with $G^+_{\rm hyb}$ in eq.~\eqref{Q4Gplus} and add the non-minimal variables $\{p_{\al 2}, \ta^{\al 2}, w_{\al}, \la^{\al}, \wb^{\al}, \lb_{\al}, s^{\al}, r_{\al} \}$, so that the supercurrent becomes
\begin{align}
{G^+}^\prime & = G^+_{\rm hyb} - \la^{\al} p_{\al 2} - \wb^{\al} r_{\al}\,.
\end{align}
Then, after performing the similarity transformation $e^{R_2} e^{R_1} {G^+}^\prime e^{-R_1} e^{-R_2} \rightarrow {G^+}^\prime$ where $R_1 = -Q^{\rm hyb}_{\al 2} \ta^{\al 2}$ and $R_2 = -\frac{i}{2} \pd x_{\al \bt} \ta^{\al 1} \ta^{\al 2}$, one learns that ${G^+}^\prime = G^+$ in \eqref{NMGplus} up to terms proportional to $\ta^{\al 2}$.\footnote{Since the BRST operator $G^+$ is supersymmetric, one can consider an additional similarity transformation to restore the missing $\ta^{\al 2}$ terms, analogously as in ref.~\cite{Berkovits:2024ono}.} This procedure is similar to the construction adopted in refs.~\cite{Berkovits:2024ono} \cite{Berkovits:2001us} in relating the RNS formalism with the pure spinor formalism. Moreover, we also learn that $e^{R_2}e^{R_1}p_{\al 2}e^{-R_1}e^{-R_2}=e^{R_2}(p_{\al 2} - Q^{\rm hyb}_{\al 2})e^{-R_2}=D_{\al}$ up to terms proportional to the non-minimal variable $\ta^{\al 2}$. The charge $Q^{\rm hyb}_{\al 2}$ was defined in eq.~\eqref{SUSYshyb}, therefore, we conclude that the constraint $D_{\al}$ is related to the ``non-standard'' SUSYs of the hybrid formalism.

Starting from the $d=10$ pure spinor formalism, there have been other approaches to describe the superstring in a six-dimensional background with manifest $d=6$ $\mathcal{N}=1$ supersymmetry \cite{Grassi:2005sb} \cite{Wyllard:2005fh} \cite{Gerigk:2009va} \cite{Chandia:2011su}. In these works, the non-minimal variables are absent but the ghosts $\{w_{\al}, \la^{\al}\}$ usually appear from the decomposition of the $d=10$ pure spinor $\la^{\overline{\al}}$, $\overline{\al} = \{1$ to $16\}$, in terms of $\rm SO(1, 5)$ spinors. Particularly, in ref.~\cite{Gerigk:2009va} a BRST operator of the form $Q_{\rm PS} = \oint \la^{\al} D_{\al}$ was proposed as the dimensional reduction of the BRST operator in the $d=10$ pure spinor formalism. In ref.~\cite{Chandia:2011su}, it was also considered adding the ghosts $\{w_{\al}, \la^{\al}\}$ to the superconformal generators \eqref{SCAgen}. The advantage of the approach detailed below is that we will be able to explicit write a BRST invariant superstring vertex operator in terms of $d=6$ $\mathcal{N}=1$ superfields and the manifest spacetime supersymmetric worldsheet variables. 

\subsection{Massless compactification-independent vertex operators}\label{secNMvertex}

We consider the compactification-independent physical states with conformal weight zero at zero momentum for the open superstring or holomorphic sector, so that we are seeking for a vertex operator $U$ which describes the $d=6$ SYM multiplet. We will start by specifying what are the physical state conditions the vertex operator has to fulfill. Then write the vertex in terms of $d=6$ $\mathcal{N}=1$ superfields depending on the eight $\ta$ coordinates. After that, it will be shown that BRST invariance of $U$ reproduces the on-shell $d=6$ SYM equations in superspace.

Since we have a nilpotent BRST charge $(G^+)_0$, we can require physical unintegrated vertex operators $U$ to be ghost-number-one states in the cohomology of $(G^+)_0$. Without loss of generality, the ghost-number current is defined to be the ${\rm U(1)}$ generator of the $\mathcal{N}=2$ algebra, eq.~\eqref{NMJ}. Moreover, the stress-tensor $T$ has vanishing conformal anomaly, it is then consistent to require $U$ to be a conformal weight zero primary field as well. When these conditions are satisfied, and given the fact that $\{(G^+)_0, (G^-)_0 \} = (T)_0$, the superconformal generator $(G^-)_0$ has to annihilate the state $U$, which means that $U$ is in the covariant Lorenz gauge \cite{Chandia:2021coc}. The latter condition is analogous to the $b_0=0$ constraint in bosonic string theory.

For the compactification-independent massless sector of the open superstring, the manifestly spacetime supersymmetric ghost-number-one unintegrated vertex operator $U$ in the cohomology of $(G^+)_0$ takes the form
\begin{align} \label{gh1vertex}
U & = - \la^{\al} \big(A_{\al 2} - A_{\al 1} e^{- \rho -i \sg} \big) + \big( \pd \ta^{\al 1} A_{\al 1} + \Pi^{\ab} A_{\ab} 
 + d_{\al 1} W^{\al 1} \big) e^{i \sg} - d_{\al 1} W^{\al 2} \pd(i \sg) e^{-\rho}    \nn \\
 &- \pd \ta^{\al 2} A_{\al 1} \pd(\rho + i\sg) e^{-\rho}+ i d_{\al 1} \Pi^{\al \bt} A_{\bt 1} e^{- \rho}  - \frac{i}{2} d_{\al 1} d_{\bt 1} A^{\al \bt} e^{-\rho}  - \pd^2 \ta^{\al 2} A_{\al 1} e^{-\rho} \nn \\
 & + \pd d_{\al 1} \big( W^{\al 2} -i \pd^{\al \bt} A_{\bt 1} \big) e^{-\rho}  + \frac{i}{2} \pd \Pi^{\al \bt} \nabla_{\al 1} A_{\bt 1} e^{-\rho}  + d_{\al 1} \big( -2 W^{\al 2} \nn \\
 &+i \pd^{\al \bt} A_{\bt 1} \big) \pd \rho e^{- \rho}  - \frac{i}{2} \Pi^{\al \bt} \nabla_{\al 1} A_{\bt 1} \pd \rho e^{-\rho} - \frac{i}{4} \pd^{\al \bt} \nabla_{\al 1} A_{\bt 1} \pd^2 e^{-\rho} + (d_1^3)^{\al} A_{\al 1} e^{- 2 \rho -i \sg} \nn \\
 &+ \ep^{\al \bt \g \dt} \bigg( - \frac{1}{4} d_{\al 1} d_{\bt 1} \nabla_{\g 1} A_{\dt 1} \pd e^{-2 \rho -i \sg} - \frac{1}{4} \pd (d_{\al 1} d_{\bt 1}) \nabla_{\g 1} A_{\dt 1} e^{-2 \rho -i \sg} \nn \\
 &- \frac{1}{12} \pd^2 d_{\al 1} \nabla_{\bt 1} \nabla_{\g 1} A_{\dt 1} e^{-2 \rho -i \sg}  - \frac{1}{6}  \pd d_{\al 1} \nabla_{\bt 1} \nabla_{\g 1} A_{\dt 1} \pd e^{-2 \rho -i \sg} \nn \\
 & - \frac{1}{12} d_{\al 1} \nabla_{\bt 1} \nabla_{\g 1} A_{\dt 1} \pd^2 e^{-2 \rho -i \sg} \bigg) + \frac{1}{4} (\nabla_1^3)^{\al} A_{\al 1} \frac{1}{6} \pd^3 e^{-2 \rho -i \sg} \,,
 \end{align}
 where $A_{\ab}$ is the superspace gauge field, $W^{\al j}$ is the superspace spinor field-strength and $F_{\ab \bb}$ is the superspace field-strength.\footnote{See Appendix \ref{sixsymapp} for a review of $d=6$ $\mathcal{N}=1$ super-Yang-Mills.} The first components of the superfields $\{A_{\ab}, W^{\al j},F_{\ab \bb}\}$ are the gluon, the gluino and the gluon field-strength, respectively. These superfields are defined in terms of the superspace gauge field $A_{\al j}$. In linearized form, we have
\begin{subequations}
\begin{align}\label{6dsuperfields}
A_{\ab} & = - \frac{i}{4} \ep^{jk} \sg_{\ab}^{\al \bt} ( \nabla_{\al j} A_{\bt k} + \nabla_{\bt k} A_{\al j} ) \,, \\
W^{\al j} & = \frac{i}{3} \ep^{jk} \sg^{\ab \al \bt}  ( \pd_{\ab} A_{\bt k}- \nabla_{\bt k} A_{\ab} ) \,, \\
F_{\ab \bb} & = \pd_{\ab} A_{\bb} - \pd_{\bb} A_{\ab}\,.
\end{align}
\end{subequations}

It is easy to see that $U$ is annihilated by $(G^-)_0$ and so we have $\pd^{\ab} A_{\ab} = 0$, which is the usual Lorenz gauge condition. The non-trivial part is showing that BRST invariance of $U$ implies the linearized $d=6$ SYM equations of motion \cite{Howe:1983fr} \cite{Koller:1982cs}
\begin{subequations}\label{gh1vertexeoms}
\begin{align}
(\sg^{\ab \bb \cb})^{\al \bt} (\nabla_{\al j} A_{\bt k} + \nabla_{\bt k} A_{\al j})&  = 0 \,, \label{gh1vertexeom1}\\
\nabla_{\al j} W^{\bt k} + \frac{i}{2} \dt^k_j (\sg_{\ab \bb})^{\bt}_{\ \al} F^{\ab \bb} & = 0\,, \label{gh1vertexeom2}
\end{align}
\end{subequations}
where $(\sg^{\ab \bb \cb})^{\al \bt} = \frac{i}{3!}(\sg^{[\ab}\sg^{\bb}\sg^{\cb]})^{\al \bt}$ is the symmetric anti-self-dual three-form and $(\sg_{\ab \bb})^{\bt}_{\ \al}= \frac{i}{2}(\sg^{[\ab}\sg^{\bb]})_{\ \al}^{\bt}$ is the generator of Lorentz transformations. 


The calculation proving that BRST invariance of $U$ leads to eqs.~\eqref{gh1vertexeoms} is detailed in Appendix \ref{newvertexapp}, and it involves taking care of various normal-ordering contributions during the process.  For example, eq.~\eqref{gh1vertexeom1} comes from the terms with $\la^{\al} \la^{\bt}$ in $(G^+)_0 U$, and eq.~\eqref{gh1vertexeom2} can be obtained by the terms proportional to $\la^{\al} d_{\bt 1}e^{i \sg}$, $\la^{\al} \pd d_{\bt 1} e^{- \rho}$, $\la^{\al} d_{\bt 1} \pd (i \sg) e^{-\rho}$ and $\la^{\al} \pd^2 d_{\bt 1} e^{-2 \rho -i \sg}$. The interested reader is referred to Appendix \ref{newvertexapp} for supplementary technical details.

Note further that $U$ in \eqref{gh1vertex} is defined up to a gauge transformation $\dt U = (G^+)_0 \La$ for some conformal weight zero and ${\rm U(1)}$-charge zero gauge parameter $\La$, and $U$ is also annihilated by $(\widetilde{G}^+_{\rm hyb})_0$ of \eqref{BVgtildeplus}, a condition that will become more clear when we write the amplitude prescription \eqref{3point2} in the following section.\footnote{When translated to the RNS variables, the condition $(\widetilde{G}^+_{\rm hyb})_0 U=0$ is equivalent as saying that $U$ lives in the small Hilbert space, i.e., it is annihilated by the $\eta_0$-ghost \cite{Friedan:1985ge}.} Taking $\La$ to be a function of the zero modes of $\{x^{\ab}, \ta^{\al j} \}$, we have that
\begin{align}
\dt U & = - \la^{\al} \big(\nabla_{\al 2} \La - \nabla_{\al 1} \La e^{- \rho -i \sg} \big) + \big( \pd \ta^{\al 1} \nabla_{\al 1} \La + \Pi^{\ab} \pd_{\ab} \La \big) e^{i \sg} + \ldots \,,
\end{align}
which precisely reproduces the gauge transformations \eqref{gaugegaugeapp} of the $d=6$ $\mathcal{N}=1$ superspace description, i.e., $\dt A_{\al j} = \nabla_{\al j} \La$ and $\dt A_{\ab} = \pd_{\ab} \La$.

For scattering amplitude computations, vertex operators in integrated form are necessary. As we have an $\mathcal{N}=2$ SCA \eqref{NMSCA}, it is straightforward to define integrated vertex operators. They are given by
\begin{align}
W = \int (G^-)_{-1} U \,,
\end{align}
which, for the compactification-independent massless sector of the open superstring, takes the simple form
\begin{align}\label{NMintvertex}
W = \int \big(\pd \ta^{\al j} A_{\al j} + \Pi^{\ab} A_{\ab} + d_{\al 1} W^{\al 1} + d_{\al 1} e^{-\rho -i \sg} W^{\al 2} \big) \,.
\end{align}
Note that only the first four terms in \eqref{gh1vertex} contribute to the integrated vertex $W$. Not surprisingly, the integrated vertex \eqref{NMintvertex} has a similar structure as in the first equality of eq.~\eqref{6dintvertexsec1}. 

The gauge transformations of $W$ are given by $\dt W = (G^+)_0 \Omega^-$ for some conformal weight one and ${\rm U(1)}$-charge minus one gauge parameter $\Omega^-$. Taking $\Omega^- = -w_{\al} W^{\al 2}$, which is annihilated by $(\widetilde{G}^+_{\rm hyb})_0$, we can write $W$ as
\begin{align}\label{NMintvertex2}
W &=  \int \bigg(\pd \ta^{\al j} A_{\al j} + \Pi^{\ab} A_{\ab} + d_{\al j} W^{\al j} - \frac{i}{2} N_{\ab \bb} F^{\ab \bb} \nn \\
& - \frac{i}{2} w_{\al} d_{\bt 1} d_{\g 1} \pd^{\bt \g} W^{\al 2} e^{-\rho} + w_{\al} \Pi^{\ab} \pd_{\ab} W^{\al 2} e^{i \sg} \bigg) \,,
\end{align}
where $N_{\ab \bb} = w_{\al} (\sg_{\ab \bb})^{\al}_{\ \bt} \la^{\al}$. 

From an argument concerning the level of the Lorentz currents in the RNS and pure spinor formalisms, the first line of \eqref{NMintvertex2} takes the form conjectured in ref.~\cite[footnote 3]{Berkovits:2000fe} to be the correct integrated vertex operator for the massless sector of the open superstring compactified to six dimensions.

\subsection{Tree-level scattering amplitudes}\label{scattamps}

In Section \ref{sec1}, we introduced an unintegrated vertex operator $V_{\rm hyb}$ with zero ${\rm U(1)}$-charge, eq.~\eqref{Vhyb}. When on-shell, this vertex operator was shown to describe $d=6$ SYM. Moreover, one can show that there exists a gauge choice where \eqref{Vhyb} can be taken to be an $\mathcal{N}=2$ superconformal primary field with respect to the SCA \eqref{BVWconstraints} \cite{Berkovits:1997zd}. 

In terms of $V_{\rm hyb}$, the tree-level three-point amplitude prescription for the massless states in the hybrid formalism of Section \ref{half6dhybrid} is \cite{Berkovits:1996bf}
\begin{align}\label{3point1}
\Big\langle V_{\rm hyb}(z_1) \big( (\widetilde{G}^+_{\rm hyb})_0 V_{\rm hyb} \big)(z_2) U_{\rm hyb}(z_3) \Big\rangle\,,
\end{align}
where $\big\langle e^{3\rho + 3i \sg} J^{++}_C (\ta)^4 \big\rangle = 1$ with $(\ta)^4 = \frac{1}{24} \ep_{\al \bt \g \dt} \ta^{\al} \ta^{\bt} \ta^{\g} \ta^{\dt}$ and we defined $U_{\rm hyb} = (G^+_{\rm hyb})_0 V_{\rm hyb}$. It is interesting to note that, in some gauge choice, $U_{\rm hyb} $ in \eqref{3point1} looks very similar to $U$ in \eqref{gh1vertex}, at least in the ghost structure when we take $\la^{\al} =0$. However, since vertex operators only depend on four $\ta$ coordinates, they do not have a simple transformation rule under all spacetime SUSYs.

We can try to use the elements of the hybrid formalism outlined in the paragraph above to formulate a prescription for calculating scattering amplitudes in terms of the superconformal generators \eqref{NMSCA} and the vertex operators in \eqref{gh1vertex} and \eqref{NMintvertex}, which are constructed from the manifestly spacetime supersymmetric variables. In this setting, recall that the eight supersymmetry generators are given by \eqref{susygen1}, as opposed to the ghost-dependent SUSYs \eqref{SUSYshyb} in the six-dimensional hybrid description.

In view of that, it is tempting to conjecture that $U$ can be written as $U= (G^+)_0 V$ for some $V$ which is also an $\mathcal{N}=2$ primary field with respect to the SCA $\eqref{NMSCA}$. Unfortunately, we could not accomplish this much and find a $V$ with both of these properties. Nonetheless, it is possible to find a conformal weight zero and ${\rm U(1)}$-charge zero field $V$ such that $U= (G^+)_0 V$ and, as we will see, this is enough to define a consistent tree-level scattering amplitude prescription. 

Consider
\begin{align}
V(z) & = \oint \frac{dy}{y-z} \big(  - (\ta^1)^4 e^{2 \rho + i \sg} \big) (y) U(z) \,,
\end{align}
and note that $(G^+)_0 V = U$ by using the fact that $(G^+)_0$ annihilates $U$ and the property
\begin{align}
(G^+)_0 \Big(- (\ta^1)^4 e^{2 \rho + i \sg} \Big)= 1\,.
\end{align}
Explicitly, the field $V$ is given by
\begin{align}\label{gh0vertex}
V & = - \la^{\al} (\ta^1)^4 A_{\al 2} e^{2 \rho + i \sg} + (\ta^1)^3_{\al} W^{\al 1} e^{2 \rho + 2 i \sg} + \bigg( \frac{i}{2} \ta^{\al 1} \ta^{\bt 1} A_{\al \bt} \nn \\
& - \frac{i}{2} (\ta^1)^4 \pd^{\al \bt} \nabla_{\al 1} A_{\bt 1} \bigg) e^{\rho + i \sg} + \ta^{\al 1} A_{\al 1} + \frac{1}{2} \ta^{\al 1} \ta^{\bt 1} \nabla_{\al 1} A_{\bt 1} \nn \\
& - \frac{1}{6} \ta^{\al 1} \ta^{\bt 1} \ta^{\g 1} \nabla_{\al 1} \nabla_{\bt 1} A_{\g 1} + \frac{1}{4} (\ta^1)^4 (\nabla^3_1)^{\al} A_{\al 1} \,,
\end{align}
where $(\ta^1)^4 = \frac{1}{24} \ep_{\al \bt \g \dt} \ta^{\al 1} \ta^{\bt 1} \ta^{\g 1} \ta^{\dt 1}$ and $(\ta^1)^3_{\al} = \frac{1}{6} \ep_{\al \bt \g \dt} \ta^{\bt 1} \ta^{\g 1} \ta^{\dt 1} $. Note that $V$ has a different ghost structure than \eqref{Vhybghoststruc}. 

In close analogy with \eqref{3point1}, the spacetime supersymmetric tree-level three-point amplitude is defined as
\begin{align}\label{3point2}
\mathcal{A}_3& =  \int [d \la] [d\lb] d^4r d^8\ta \, \mathcal{R} \Big\langle V(z_1) \big( (\widetilde{G}^+_{\rm hyb})_0 V \big)(z_2) U(z_3) \Big\rangle \,,
\end{align}
where $V$ is given by \eqref{gh0vertex}, $G^+_{\rm hyb}$ is given by \eqref{BVgtildeplus} and $U=(G^+)_0V$ is the ghost number one vertex operator in eq.~\eqref{gh1vertex}. We also define $\big\langle e^{3\rho + 3i \sg} J^{++}_C \big\rangle = 1$, due to the anomaly in the ${\rm U(1)}$ current. 

Since the bosonic variables $\la^{\al}$ and $\lb_{\al}$ are non-compact, a regularization factor $\mathcal{R}= \exp ((G^+)_0 \chi)$ needs to be introduced. We will take $\chi = \lb_{\al} \ta^{\al 2}$ \cite{Berkovits:2005bt}, so that one finds
\begin{align}
\mathcal{R} = \exp \big( -\la^{\al} \lb_{\al} + r_{\al} \ta^{\al 2} \big) \,. 
\end{align}
For simplicity, the integration over the $x^{\ab}$ zero modes is being ignored, since it is done in the standard manner \cite{Polchinski:1998rq}. Given that the expression inside brackets is BRST invariant and $\mathcal{R} = 1 + (G^+)_0 (\ldots)$, the amplitude \eqref{3point2} is independent of $\chi$ as long as $\chi$ is annihilated by $(G^+_{\rm hyb})_0$.

Despite the asymmetric appearance, the amplitude \eqref{3point2} is symmetric in the three insertions. This is easy to see by noting that $(\widetilde{G}^+_{\rm hyb})_0 U = (\widetilde{G}^+_{\rm hyb})_0 \chi = 0$ and $\{(\widetilde{G}^+_{\rm hyb})_0, (G^+)_0 \} = 0$. As long as one chooses $\chi$ such that $(\widetilde{G}^+_{\rm hyb})_0 \chi=0$, the amplitude \eqref{3point2} will be independent of the choice of $\chi$. Since the $(\widetilde{G}^+_{\rm hyb})_0$ cohomology is trivial one can even choose $\chi$ to be exact.

From the $c=6$ $\mathcal{N}=2$ SCA \eqref{NMSCA}, it is straightforward to use the procedure outlined in Appendix \ref{SCAs} and construct the remaining generators of the small $c=6$ $\mathcal{N}=4$ SCA. In such a case, one could have thought that it would be possible to define the amplitude \eqref{3point2} with the superconformal generator $\widetilde{G}^+$ of the $\mathcal{N}=4$ algebra associated with $\eqref{NMSCA}$ instead of $\widetilde{G}^+_{\rm hyb}$ in \eqref{BVgtildeplus}. However, it turns out that an amplitude defined in this way would give a vanishing result. The reason for this is that $\widetilde{G}^+$ involves an overall factor containing $\dt^4(r)$,\footnote{This is easier to see in the bosonized form of $\{w_{\al}, \la^{\al}, s^{\al}, r_{\al}\}$.} but we already have the four zero modes of $r_{\al}$ and $\ta^{\al 2}$ coming from the regulator $\mathcal{R}$. The issue arising when trying to use $\widetilde{G}^+$ in our prescription might be related to the fact that physical states of the superstring cannot be defined as $\mathcal{N}=2$ primaries with respect to the algebra \eqref{NMSCA}.

The amplitude \eqref{3point2} is gauge-invariant under $\dt V =  (G^+)_0 \Lambda + (G^+_{\rm hyb})_0 \Omega$. Since $U$ satisfies $(G^+_{\rm hyb})_0 U =0$ and $U=(G^+)_0V$, we have that $V$ obeys the equation $(G^+_{\rm hyb})_0 (G^+)_0 V =0 $, which is invariant under the gauge transformation $\dt V =  (G^+)_0 \Lambda + (G^+_{\rm hyb})_0 \Omega$ for any $\{ \Omega, \Lambda\}$. 

The amplitude \eqref{3point2} is supersymmetric. Although the regulator is not manifestly spacetime supersymmetric, its spacetime supersymmetry transformation under the generators \eqref{susygen1} is BRST trivial, and hence vanishes inside the amplitude expression \eqref{3point2}. Moreover, the vertex operator $U$ is written in terms of the supersymmetric worldsheet variables, and we have shown that the amplitude is symmetric in the three insertions.

As a result, we have verified that the scattering amplitude prescription \eqref{3point2} for the six-dimensional compactification-independent states satisfies all the desired properties: (i) it is gauge-invariant; (ii) it is $d=6$ $\mathcal{N}=1$ spacetime supersymmetric; (iii) it reduces to the $\mathcal{N}=4$ topological prescription \eqref{3point1} of the hybrid formalism \cite{Berkovits:1994vy} when the additional variables $\{p_{\al 2}, \ta^{\al 2}, w_{\al}, \la^{\al}, r_{\al}, s^{\al}, \wb^{\al}, \lb_{\al} \}$ are set to zero; and (iv) it is symmetric in the vertex operator insertions.

In order to further check the consistency of our proposal, let us compute the three-point amplitude involving three massless states \eqref{gh0vertex}. To simplify the analysis, we will consider the three gluon amplitude $\mathcal{A}_{BBB}$, so that we can effectively put the gluino to zero in the $d=6$ SYM superfields (see eqs.~\eqref{thetaexpSYM}). In this particular case, we have that $(\ta^1)^3_{\al} W^{\al 1} =0$ in \eqref{gh0vertex}. Furthermore, the non-zero contributions to \eqref{3point2} can be determined by looking at which terms have the right amount of ghost insertions to saturate the background charge of the $\{ \rho ,\sg \}$ ghosts, we are then left with the following worldsheet correlator
\begin{align} \label{ampnew}
\mathcal{A}_{BBB}& =  \int [d \la] [d\lb] d^4r d^8\ta \, \mathcal{R} \bigg\langle \bigg( \frac{i}{2} \ta^{\al 1} \ta^{\bt 1} A^{(1)}_{\al \bt} e^{\rho + i \sg} \bigg) (z_1) \times \nn \\
& \times \bigg( \frac{i}{2} \ta^{\g 1} \ta^{\dt 1} A^{(2)}_{\g \dt} e^{2 \rho + i \sg}J^{++}_C \bigg)(z_2) \Big( \Pi^{\ab} A^{(3)}_{\ab} + d_{\sg 1} W^{(3)\sg 1} \Big) e^{i \sg} (z_3) \bigg\rangle + (2 \leftrightarrow 3) \,,
\end{align}
and, after using $\rm{SL(2,\mathbb{R})}$ invariance to choose $z_1= \infty$, $z_2 = 1 $ and $z_3 = 0$, it easy to see that
\begin{align}
\mathcal{A}_{BBB}& = -i \big( (a_1 \cdot a_2 ) (k_2 \cdot a_3) + (a_1 \cdot a_3 ) (k_1 \cdot a_2) + (a_2 \cdot a_3) (k_3 \cdot a_1) \big) + (2 \leftrightarrow 3) \,,
\end{align}
which gives the sought after result, as expected. Since $U$ describes the $d=6$ SYM multiplet, and by invariance under $d=6$ $\mathcal{N}=1$ supersymmetry transformations, we can conclude that our prescription also reproduces the expected answer for the three-point amplitude involving one gluon and two gluinos $\mathcal{A}_{BFF}$.

It is then elementary to generalize \eqref{3point2} to the case where we have $n$ super-Yang-Mills multiplets
\begin{align}\label{npointexthyb}
\mathcal{A}_n& =  \int [d \la] [d\lb] d^4r d^8\ta \, \mathcal{R} \bigg\langle V(z_1) \big( (\widetilde{G}^+_{\rm hyb})_0 V \big)(z_2) U(z_3) \prod_{m=4}^n \int dz_m  (G^-)_{-1} U(z_m)  \bigg\rangle \,,
\end{align}
where $\{z_1, z_2 ,z_3\}$ can be chosen arbitrarily by $\rm SL(2, \mathbb{R})$ invariance. As we have only described vertex operators for the massless compactification-independent states, just scattering of $d=6$ SYM multiplets was considered, however, the tree-level prescription should also apply to massive compactification-independent states.

\section{Conclusion} \label{concl}

In this work, we have studied the superstring compactified to a six-dimensional background and its description with manifest $d=6$ $\mathcal{N}=1$ supersymmetry. After relaxing the harmonic first-class constraint $D_{\al}$ of ref.~\cite{Berkovits:1999du} and defining a new BRST operator $G^+$, spacetime supersymmetric vertex operators and a tree-level scattering amplitude prescription were constructed. Specifically, it was shown that BRST invariance of the vertex operator imply the $d=6$ SYM equations of motion in $\mathcal{N}=1$ superspace. Furthermore, we confirmed that the three-point amplitude of SYM states is reproduced.

An immediate application of this work would be to generalize the tree-level prescription \eqref{npointexthyb} to a multiloop amplitude prescription for an arbitrary number of $d=6$ SYM multiplets. In this case, the regulator $\mathcal{R}$ should include the genus $g$ zero-modes \cite{Berkovits:2005bt} of the worldsheet fields, and the prescription \eqref{npointexthyb} should be modified, possibly with insertions of the $\mathcal{N}=2$ generators in a genus-$g$ surface \cite{Berkovits:1994vy}. Since the $b$-ghost (or $G^-$ in our case) does not have singularities when $\{\la^{\al}, \lb_{\al}\} \rightarrow 0$, there is no restriction in the number of $b$-ghost insertions as in the non-minimal pure spinor case \cite{Berkovits:2005bt}.

Let us point out that the term $-\la^{\al} D_{\al}$ in the BRST operator \eqref{NMGplus} has been proposed as the BRST operator of the six-dimensional pure spinor formalism in refs.~\cite{Wyllard:2005fh} \cite{Gerigk:2009va}, where a six-dimensional pure spinor $\la^{\al j}$ is defined to satisfy $\ep_{jk}\la^{\al j} \sg^{\ab}_{\al \bt} \la^{\bt k}=0$, so that it has five independent components.  In this case, it was shown that $\la^{\al j}$ is related to the hybrid formalism $\{\rho, \sg\}$-ghosts and the bosonic ghost $\la^{\al}$ as $\la^{\al j}= \{-e^{-\rho -i \sg}\la^{\al}, \la^{\al}\}$ \cite{Gerigk:2009va}. Therefore, the term $-\la^{\al} D_{\al}$ can be written in a way that the outer $\rm SU(2)$ symmetry is manifst, i.e., as $-\la^{\al j} d_{\al j}$ for $\la^{\al j}$ a six-dimensional pure spinor, where $d_{\al j}$ is defined in eqs.~\eqref{susyvars}. 

However, as opposed to the $d=10$ pure spinor formalism \cite{Berkovits:2000fe}, defining $Q_{\rm PS}= \oint \la^{\al j} d_{\al j}$ to be the six-dimensional BRST operator is not enough to imply on-shell $d=6$ SYM from the naive vertex operator $\la^{\al j} A_{\al j}$.\footnote{More precisely, eq.~\eqref{gh1vertexeom2} is not reproduced by invariance of $\la^{\al j} A_{\al j}$ under $Q_{\rm PS}=\oint \la^{\al j} d_{\al j}$.} It would be interesting to further study this connection between the extended hybrid formalism presented in this paper and a possible six-dimensional pure spinor description of the superstring involving the constrained $\la^{\al j}$. The latter might be a promising research direction for uncovering a description of the superstring in $d=6$ $\mathcal{N}=1$ harmonic superspace \cite{Howe:1985ar}.

\section*{Acknowledgements}

CAD would like to thank Nathan Berkovits for discussions and suggestions, as well as FAPESP grant numbers 2022/14599-0 and 2023/00015-0 for financial support.


\appendix

\section{\boldmath $\mathcal{N}=2$ and small $\mathcal{N}=4$ superconformal algebras} \label{SCAs}

We present the general structure of $\mathcal{N}=2$ and small $\mathcal{N}=4$ superconformal algebras, as well as their twisted counterparts. We do not try to address questions such as when and how these algebras can be realized.

\subsection{$\mathcal{N}=2$ superconformal algebra} \label{sca}

The $\mathcal{N}=2$ superconformal algebra with central charge $c$ satisfied by the generators $\{J,G^+,G^-,T\}$ is given by
\begin{subequations}\label{SCAapp}
\begin{align}
T(y)T(z) & \sim \frac{\frac{c}{2}}{(y-z)^4} + \frac{2 T(z)}{(y-z)^2} +\frac{ \pd T(z)}{(y-z)} \,, \\
G^+(y)G^-(z) & \sim \frac{\frac{c}{3}}{(y-z)^3} + \frac{J(z)}{(y-z)^2} + \frac{T(z) + \frac{1}{2} \pd J(z)}{(y-z)}\,, \\
T(y)G^{\pm}(z) & \sim \frac{\frac{3}{2} G^{\pm}(z)}{(y-z)^2} + \frac{\pd G^{\pm}(z)}{(y-z)}\,, \\
T(y)J(z) & \sim \frac{J(z)}{(y-z)^2} + \frac{ \pd J(z)}{(y-z)}\,, \\
J(y)J(z) & \sim \frac{\frac{c}{3}}{(y-z)^2}\,, \\
J(y)G^{\pm}(z)& \sim \pm \frac{G^{\pm}(z)}{(y-z)}\,.
\end{align}
\end{subequations} 
Here, $T$ has conformal weight 2, $G^{\pm}$ has conformal weight $\frac{3}{2}$ and $J$ has conformal weight 1.

Equivalently, in tems of the modes, the $\mathcal{N}=2$ SCA reads
\begin{subequations}
\begin{align}
[L_m, L_n] & = (m-n) L_{m+n} + \tfrac{c}{12}(m^3 -m) \dt_{m,-n}\,, \\
\{G^+_r, G^-_s \} & = L_{r+s} + \tfrac{1}{2} (r-s) J_{r+s} + \tfrac{c}{6} (r^2 - \tfrac{1}{4}) \dt_{r,-s} \,, \\
[L_m, G^{\pm}_r] & = ( \tfrac{1}{2}m -r) G^{\pm}_{m+r}\,, \\
[L_m, J_n] & = -n J_{m+n}\,, \\
[J_m, J_n] & = \tfrac{c}{3} m \dt_{m , -n}\,, \\
[J_m, G^{\pm}_r] & = \pm G^{\pm}_{m+r}\,.
\end{align}
\end{subequations}

\subsection{Twisted $\mathcal{N}=2$ superconformal algebra}

To construct an $\mathcal{N}=2$ twisted theory, we modify the stress-tensor $T$ by adding $+\frac{1}{2} \pd J$ to it, so that
\begin{equation}
T + \frac{1}{2} \pd J \mapsto T\,, \label{stressshift}
\end{equation}
and one can see that the dimension of every field in the theory is modified by $-\frac{1}{2}$ its U(1)-charge, which is generated by $J$. In particular, looking at the structure of the algebra \eqref{SCAapp}, we see that the conformal weight of $G^+$ gets shifted to 1, that of $G^-$ gets shifted to 2 and the conformal weight of the rest of the generators stay untouched. More importantly, the shift in the stress-tensor \eqref{stressshift} results in the vanishing of the conformal anomaly in the $TT$ OPE, so that the twisted stress-tensor is a primary. In contrast, there appears a triple pole in the $TJ$ OPE proportional to the central charge $c$.

With the above considerations, we can now write the twisted $\mathcal{N}=2$ superconformal algebra with central charge $c$ satisfied by the twisted generators $\{J,G^+,G^-,T\}$\footnote{Here, $T$ is the shifted stress-tensor of \eqref{stressshift}.}
\begin{subequations} \label{SCAapptw}
\begin{align}
T(y)T(z) & \sim \frac{2 T(z)}{(y-z)^2} +\frac{ \pd T(z)}{(y-z)} \,, \\
G^+(y)G^-(z) & \sim \frac{\frac{c}{3}}{(y-z)^3} + \frac{J(z)}{(y-z)^2} + \frac{T(z)}{(y-z)}\,, \\
T(y)G^{+}(z) & \sim \frac{G^{+}(z)}{(y-z)^2} + \frac{\pd G^{+}(z)}{(y-z)}\,, \\
T(y)G^{-}(z) & \sim \frac{2G^{-}(z)}{(y-z)^2} + \frac{\pd G^{-}(z)}{(y-z)}\,, \\
T(y)J(z) & \sim - \frac{\frac{c}{3}}{(y-z)^{3}} + \frac{J(z)}{(y-z)^2} + \frac{ \pd J(z)}{(y-z)}\,, \\
J(y)J(z) & \sim \frac{\frac{c}{3}}{(y-z)^2}\,, \\
J(y)G^{\pm}(z)& \sim \pm \frac{G^{\pm}(z)}{(y-z)}\,.
\end{align}
\end{subequations}%

\subsection{Small and twisted small $\mathcal{N}=4$ superconformal algebras}

A small $\mathcal{N}=4$ superconformal algebra consists of a conformal weight 2 generator $T$, four conformal weight $\frac{3}{2}$ fermionic currents $\{G^{\pm}, \widetilde{G}^{\pm}\}$ and three conformal weight 1 bosonic currents $\{J,J^{++},J^{--}\}$ forming an $\mathfrak{su}(2)_{\frac{c}{6}}$ current algebra. In the description that we are using, it is convenient to build the small $\mathcal{N}=4$ SCA by starting with the $\mathcal{N}=2$ SCA in Appendix \eqref{sca} and lifting the  $\mathfrak{u}(1)_{\frac{c}{6}}$ to an $\mathfrak{su}(2)_{\frac{c}{6}}$ current algebra. To do that, one adds to the generators $\{J,G^+,G^-,T\}$ the conformal weight 1 bosonic currents $J^{++}$ and $J^{--}$ of ${\rm U(1)}$ charge $\pm 2$, respectively, satisfying the OPES
\begin{subequations} \label{su(2)curr}
\begin{align}
J(y)J^{\pm \pm}(z) & \sim \pm 2  \frac{J^{\pm \pm}(z)}{(y-z)}\,, \\
J^{++}(y)J^{--}(z) & \sim \frac{\frac{c}{6}}{(y-z)^{2}} + \frac{J(z)}{(y-z)}\,. 
\end{align}
\end{subequations}

Note that the level of the $\mathfrak{su}(2)$ current algebra is fixed by the Jacobi identities and the level of the $\mathfrak{u}(1)$ current algebra. On top of that, for the algebra to close, we also need to add two fermionic generators $\widetilde{G}^{\pm}$ and, in addition to the non-regular OPEs in eq.~\eqref{SCAapp}, we also have
\begin{subequations}\label{SCAapp2}
\begin{align}
J^{\pm \pm}(y) G^{\mp}(z) & \sim \mp \frac{\widetilde{G}^{\pm}(z)}{(y-z)}\,, \\
J^{\pm \pm}(y) \widetilde{G}^{\mp}(z) & \sim \pm \frac{G^{\pm}(z)}{(y-z)}\,, \\
G^{+}(y) \widetilde{G}^{+}(z) & \sim \frac{2 J^{++}(z)}{(y-z)^2} + \frac{\pd J^{++}
(z)}{(y-z)}\,, \\
\widetilde{G}^{-}(y) G^{-}(z) & \sim \frac{2 J^{--}(z)}{(y-z)^2} + \frac{\pd J^{--}(z)}{(y-z)}\,, \\
\widetilde{G}^+(y)\widetilde{G}^-(z) & \sim \frac{\frac{c}{3}}{(y-z)^3} + \frac{J(z)}{(y-z)^2} + \frac{T(z) + \frac{1}{2} \pd J(z)}{(y-z)}\,, \\
T(y)J^{\pm \pm}(z) & \sim \frac{J^{\pm \pm}(z)}{(y-z)^{2}} + \frac{ \pd J^{\pm \pm}(z)}{(y-z)}\,, \\
T(y)\widetilde{G}^{\pm}(z) & \sim \frac{\frac{3}{2} \widetilde{G}^{\pm}(z)}{(y-z)^2} + \frac{\pd \widetilde{G}^{\pm}(z)}{(y-z)}\,.
\end{align}%
\end{subequations}%
Therefore, we say that the generators $\{J, J^{\pm \pm}, G^{\pm}, \widetilde{G}^{\pm}, T\}$ form a small $\mathcal{N}=4$ SCA with central charge $c$ when they satisfy eqs.~\eqref{SCAapp}, \eqref{su(2)curr} and \eqref{SCAapp2}.

The twisted small $\mathcal{N}=4$ SCA with central charge $c$ can be constructed from the untwisted one in the same way as we constructed the twisted $\mathcal{N}=2$ SCA from eq.~\eqref{SCAapp}, i.e., by shifting the stress-tensor as in eq.~\eqref{stressshift}. With respect to the twisted stress-tensor, the conformal weight of $J^{++}$ becomes zero, that of $J^{--}$ becomes 2, the conformal weight of $G^+$ and $\widetilde{G}^+$ gets shifted to 1 and that of $G^{-}$ and $\widetilde{G}^-$ gets shifted to 2. Consequently, we say that the twisted generators $\{J, J^{\pm \pm}, G^{\pm}, \widetilde{G}^{\pm}, T\}$ form a twisted small $\mathcal{N}=4$ SCA with central charge $c$ when they obey eqs.~\eqref{SCAapptw}, \eqref{su(2)curr} and
\begin{subequations}
\begin{align}
J^{\pm \pm}(y) G^{\mp}(z) & \sim \mp \frac{\widetilde{G}^{\pm}(z)}{(y-z)}\,, \\
J^{\pm \pm}(y) \widetilde{G}^{\mp}(z) & \sim \pm \frac{G^{\pm}(z)}{(y-z)}\,, \\
G^{+}(y) \widetilde{G}^{+}(z) & \sim \frac{2 J^{++}(z)}{(y-z)^2} + \frac{\pd J^{++}
(z)}{(y-z)}\,, \\
\widetilde{G}^{-}(y) G^{-}(z) & \sim \frac{2 J^{--}(z)}{(y-z)^2} + \frac{\pd J^{--}(z)}{(y-z)}\,, \\
\widetilde{G}^+(y)\widetilde{G}^-(z) & \sim \frac{\frac{c}{3}}{(y-z)^3} + \frac{J(z)}{(y-z)^2} + \frac{T(z) }{(y-z)}\,, \\
T(y)J^{++}(z) & \sim \frac{ \pd J^{++}(z)}{(y-z)}\,, \\
T(y)J^{--}(z) & \sim \frac{2J^{--}(z)}{(y-z)^{2}} + \frac{ \pd J^{--}(z)}{(y-z)}\,, \\
T(y)\widetilde{G}^{+}(z) & \sim \frac{\widetilde{G}^{+}(z)}{(y-z)^2} + \frac{\pd \widetilde{G}^{+}(z)}{(y-z)}\,, \\
T(y)\widetilde{G}^{-}(z) & \sim \frac{2\widetilde{G}^{-}(z)}{(y-z)^2} + \frac{\pd \widetilde{G}^{-}(z)}{(y-z)}\,.
\end{align}
\end{subequations}

With respect to the $\mathfrak{su}(2)$ symmetry, $T$ transforms as a singlet and $G^+$ $(G^-)$ transforms as an upper (lower) component of an $\mathfrak{su}(2)$ doublet whose lower (upper) component is $\widetilde{G}^-$($\widetilde{G}^+)$. This $\mathfrak{su}(2)$ rotates the different choices of the ${\rm U(1)}$ current $J$ into one another and computations are equivalent no matter what choice of this ${\rm U(1)}$ one picks \cite{Berkovits:1994vy}. 

In addition, there is another SU(2) symmetry (that we refer to as SU(2)$_{outer}$) of the $\mathcal{N}=4$ SCA which acts by outer automorphisms. To see that, consider the following linear combinations of the fermionic generators\footnote{Note that here they obey the hermiticity properties $(\mathbf{G}^\pm)^* = \mathbf{\widetilde{G}}^{\mp}$ and $(\mathbf{\widetilde{G}}^\pm)^*=\mathbf{\widetilde{G}}^\mp$.}
\begin{subequations}
\begin{align}
\mathbf{G}^+ & = u_{1}^* G^+ - u_{2}^* \widetilde{G}^+\,, \\
\mathbf{G}^- & = u_{1} G^- - u_2 \widetilde{G}^-\,, \\
\mathbf{\widetilde{G}}^+ & = u_1 \widetilde{G}^+ + u_2 G^+\,, \\
\mathbf{\widetilde{G}}^- & = u_1^* \widetilde{G}^- + u_2^* G^-\,,
\end{align}
\end{subequations}
by demanding that $\mathbf{G}^{\pm}$ and $\mathbf{\widetilde{G}}^\pm$ satisfy the same algebra as $G^\pm$ and $\widetilde{G}^\pm$ we get the relation $|u_1|^2 + |u_2|^2 = 1$, i.e., $u_1$ and $u_2$ are elements of SU(2)$_{outer}$. This symmetry that rotates the supercurrents parametrizes the different embeddings of the $\mathcal{N}=2$ SCA into the $\mathcal{N}=4$ SCA and, in general, is not a symmetry of the theory \cite{Berkovits:1994vy} \cite{Berkovits:1999im}.
 
Lastly, we should mention the important fact that a small $\mathcal{N}=4$ SCA can be constructed from any $c=6$ $\mathcal{N}=2$ SCA by defining the $SU(2)$ currents to be $J$, $J^{++}=-e^{\int J}$ and $J^{--}= e^{-\int J}$. The condition $c=6$ is necessary in order for $J^{++}$ and $J^{--}$ to have conformal weight 1 when the algebra is not twisted. As an example, the RNS superstring has a description as a $c=6$ $\mathcal{N}=2$ string and, therefore, can also be described as an $\mathcal{N}=4$ topological string \cite{Berkovits:1994vy}.

\section{Normal-ordering prescription}\label{normalO}

The normal-ordered product of the operators $\mathcal{O}_1$ and $\mathcal{O}_2$ is denoted by $(\mathcal{O}_1\mathcal{O}_2)$, which is defined as
\begin{equation}
(\mathcal{O}_1\mathcal{O}_2)(z) = \oint \frac{dx}{x-z} \mathcal{O}_1(x) \mathcal{O}_2(z)\,. \label{defnord}
\end{equation}
This prescription consists in subtracting the poles evaluated at the point of the second entry. By convention, when nothing is specified, our expressions are normal-ordered from the right, e.g., $\mathcal{O}_1\mathcal{O}_2\mathcal{O}_3...\mathcal{O}_n=(\mathcal{O}_1(\mathcal{O}_2(\mathcal{O}_3(...\mathcal{O}_n)...)))$.
Also, whenever we are dealing with derivatives of exponentials, such as $\pd^2 e^{\rho}$, the ordering is always done with the exponential on the right, so that $\pd^2 e^{\rho}=(\pd\rho((\pd \rho) e^{\rho}))+((\pd^2\rho) e^{\rho})$. Putting the exponential on the rightmost position agrees with the usual conformal-normal-ordering \cite{Polchinski:1998rq} when dealing with free fields. 

Schematically, note that in terms of the definition in eq.~\eqref{defnord}, we have \cite{DiFrancesco:1997nk}
\begin{equation}
(\mathcal{O}_1(\mathcal{O}_2\mathcal{O}_3))(z)= \oint \frac{dx}{x-z}\,\mathcal{O}_1(x) (\mathcal{O}_2\mathcal{O}_3)(z) = \oint \frac{dx}{x-z}\oint \frac{dy}{y-z}\, \mathcal{O}_1(x)\mathcal{O}_2(y)\mathcal{O}_3(z)\,.
\end{equation}

\section{\boldmath Determining $G^+$} \label{gpluscalc}

\subsection{Details of the computation}
We have that
\begin{equation}
G^+_{\rm hyb} =-\frac{1}{24} \ep^{\al \bt \g \dt} [D_{\al},\{ D_{\bt},[ D_{\g},\{ D_{\dt},e^{2 \rho +3 i\sg}\}]\}] + G_C^+\,, \label{gplus}
\end{equation}
where $D_{\al}= d_{\al 2} - e^{-\rho -i\sg} d_{\al1}$ and the graded bracket $[D_{\al}, \mathcal{O} \}$ denotes the single pole in the OPE between $D_{\al}$ and $\mathcal{O}$. In the following, we evaluate each of the four contributions separately.
\paragraph{First contribution.}
\begin{align}
\oint dy\, D_{\dt}(y) e^{2 \rho + 3i \sg}(z)&= \oint dy\, (d_{\dt 2} - d_{\dt 1} e^{-\rho -i \sg})(y)e^{2 \rho + 3i \sg}(z)\nn\\
&= - (d_{\dt1} e^{\rho +2i\sg})(z)\,. \label{first}
\end{align}
The term appearing in \eqref{first} comes from the single pole in the OPE between \\ $(d_{\dt 1} e^{-\rho -i \sg})(y)$ and $e^{2 \rho + 3i \sg}(z)$.
\paragraph{Second contribution.}
\begin{align}
-\oint dy\, D_{\g}(y)(d_{\dt1} e^{\rho +2i\sg})(z)&= -\oint dy\, (d_{\g2} - d_{\g 1}e^{-\rho -i \sg})(y)(d_{\dt1} e^{\rho +2i\sg})(z) \nn \\
&= i(\Pi_{\g \dt} e^{\rho +2i \sg})(z) - (d_{\g 1} d_{\dt 1} e^{i\sg})(z)\,. \label{second}
\end{align}
The first term in \eqref{second} comes from the single pole in the OPE of $d_{\g2}(y)$ and \\$(d_{\dt1} e^{\rho +2i\sg})(z)$. The second term comes from the single pole in the OPE between $(d_{\g 1}e^{-\rho -i \sg})(y)$ and $(d_{\dt1} e^{\rho +2i\sg})(z)$.
\paragraph{Third contribution.}
Now we need to compute $\oint dy\, D_{\bt}(y)\Big(i(\Pi_{\g \dt} e^{\rho +2i \sg})(z)$ \\$ - (d_{\g 1} d_{\dt 1} e^{i\sg})(z)\Big)$, which is most easily obtained by calculating the relevant terms independently. We have that
\begin{align}
&\oint dy\, d_{\bt2}(y) (-1)(d_{\g 1} d_{\dt 1} e^{i\sg})(z)\nn \\ 
&\qquad=-\oint dy\,d_{\bt 2}(y) \oint \frac{dx}{x-z}\,d_{\g 1}(x)(d_{\dt 1} e^{i\sg})(z)\nn\\
&\qquad=-\oint dy \oint \frac{dx}{x-z}\,\Big[-i (y-x)^{-1} \Pi_{\bt \g}(x)(d_{\dt 1} e^{i\sg})(z) \nn\\
& \qquad  - d_{\g 1}(x) \Big(-i(y-z)^{-1} (\Pi_{\bt \dt}e^{i \sg})(z) \Big) \Big] \nn \\
&\qquad= i (\Pi_{\bt \g}(d_{\dt 1} e^{i \sg}))(z) - i (d_{\g 1}(\Pi_{\bt \dt} e^{i \sg}))(z) \nn \\
&\qquad=i(d_{\dt 1}(\Pi_{\bt \g} e^{i \sg}))(z) - i(d_{\g 1}(\Pi_{\bt \dt} e^{i \sg}))(z) + \ep_{\ep \bt \g \dt} (\pd^2 \ta^{\ep 2} e^{i \sg})(z)\,, \label{th1}
\end{align}
where we used that $([\Pi_{\bt \g},d_{\dt1}])=\oint dy\, \Big(\Pi_{\bt \g}(y) d_{\dt1}(z) - d_{\dt 1}(y) \Pi_{\bt \g}(z)\Big) =$  \\$-i \ep_{\ep \bt \g \dt} \pd^2 \ta^{\ep 2}(z)$ according to eqs.~\eqref{dPi}. We also need
\begin{align}
\oint dy \, d_{\bt 2}(y) i(\Pi_{\g \dt} e^{\rho +2i \sg})(z) &= \ep_{\ep \bt \g \dt}(\pd \ta^{\ep 1} e^{\rho + 2 i \sg})(z)\,. \label{th2}
\end{align}
And
\begin{align}
\oint dy \, (d_{\bt 1} e^{-\rho -i \sg})(y)(d_{\g 1}d_{\dt 1} e^{i\sg})(z)=(d_{\bt 1} d_{\g 1} d_{\dt 1} e^{-\rho})(z)\,. \label{th3}
\end{align}
And lastly, we have
\begin{align}
&\oint dy \, (-i) (d_{\bt 1} e^{-\rho -i\sg})(y)(\Pi_{\g \dt} e^{\rho + 2i \sg})(z) \nn\\
&\qquad= -i \oint dy\,(d_{\bt 1} e^{-\rho -i\sg})(y)\oint \frac{dx}{x-z}\, \Pi_{\g \dt}(x) e^{\rho + 2i \sg}(z) \nn \\
&\qquad= -i\oint dy \oint \frac{dx}{x-z}\,\Big( i (y-x)^{-1} \ep_{\ep \bt \g \dt}(\pd \ta^{\ep 2} e^{-\rho -i\sg})(x)e^{\rho + 2i \sg}(z)\nn \\
 & \qquad - \Pi_{\g \dt}(x) (y-z)^{-1}(d_{\bt 1}e^{i \sg})(z)\Big) \nn \\
 & \qquad=\ep_{\ep \bt \g \dt}((\pd \ta^{\ep 2} e^{-\rho -i \sg})e^{\rho + 2i \sg})(z) + i(\Pi_{\g \dt}(d_{\bt 1} e^{i\sg}))(z) \nn \\
 & \qquad=\ep_{\ep \bt \g \dt}(\pd \ta^{\ep 2}(\pd(\rho + i \sg)e^{i\sg}))(z) + i (d_{\bt 1}(\Pi_{\g \dt} e^{i \sg}))(z)\,, \label{th4}
\end{align}
where it was used that $\ep_{\ep \bt \g \dt}((\pd \ta^{\ep 2} e^{-\rho -i \sg})e^{\rho + 2i \sg})=\ep_{\ep \bt \g \dt}(\pd \ta^{\ep 2}(\pd(\rho + i \sg)e^{i\sg}))$ \\ $ -\ep_{\ep \bt \g \dt}(\pd^2 \ta^{\ep 2} e^{i \sg}) $ and $i(\Pi_{\g \dt}(d_{\bt 1} e^{i\sg})) = i (d_{\bt 1}(\Pi_{\g \dt} e^{i \sg}))+\ep_{\ep \bt \g \dt}(\pd^2 \ta^{\ep 2} e^{i \sg})$ to go from the third to the last line in the computation of \eqref{th4}.

Gathering eqs.~\eqref{th1}--\eqref{th4}, we have 
\begin{align}
&\oint dy\, D_{\bt}(y)\Big(i(\Pi_{\g \dt} e^{\rho +2i \sg})(z) - (d_{\g 1} d_{\dt 1} e^{i\sg})(z)\Big) \nn\\
&\qquad= (d_{\bt 1} d_{\g 1} d_{\dt 1} e^{-\rho})(z) + \ep_{\ep \bt \g \dt}(\pd \ta^{\ep 1} e^{\rho + 2 i \sg})(z) + \ep_{\ep \bt \g \dt} (\pd^2 \ta^{\ep 2} e^{i \sg})(z) \nn\\
&\qquad + \ep_{\ep \bt \g \dt}(\pd \ta^{\ep 2}(\pd(\rho + i \sg)e^{i\sg}))(z) + i (d_{\bt 1}(\Pi_{\g \dt} e^{i \sg}))(z) + i(d_{\dt 1}(\Pi_{\bt \g} e^{i \sg}))(z) \nn\\
& \qquad - i(d_{\g 1}(\Pi_{\bt \dt} e^{i \sg}))(z)\,. \label{th}
\end{align}
\paragraph{Fourth contribution.} According to eq.~\eqref{gplus}, to obtain $G^+_{\rm hyb}$, we still need to act with $-\frac{1}{24}\ep^{\al \bt \g \dt} \oint dy \, D_{\al}(y)$ in eq.~\eqref{th}. We get that
\begin{align}
&-\frac{1}{24}\ep^{\al \bt \g \dt} \oint dy \, D_{\al}(y)(d_{\bt 1} d_{\g 1} d_{\dt 1} e^{-\rho})(z) \nn \\
& \qquad= -\frac{1}{24}\ep^{\al \bt \g \dt} \oint dy\,\Big(d_{\al 2}(y)\oint \frac{dx}{x-z}\, d_{\bt 1}(x) (d_{\g 1} d_{\dt 1} e^{-\rho})(z) \nn \\
& \qquad- (d_{\al 1} e^{-\rho -i\sg})(y) (d_{\bt 1} d_{\g 1} d_{\dt 1} e^{-\rho})(z) \Big) \nn \\
&\qquad=-\frac{1}{24}\ep^{\al \bt \g \dt} \oint dy \oint \frac{dx}{x-z}\, \Big[-i (y-x)^{-1} \Pi_{\al \bt}(x) (d_{\g 1} d_{\dt 1} e^{-\rho})(z) \nn \\
& \qquad- d_{\bt 1}(x) \Big( -i (y-z)^{-1} (\Pi_{\al \g}(d_{\dt1}e^{-\rho}))(z) + i (y-z)^{-1}(d_{\g 1}(\Pi_{\al \dt} e^{-\rho}))(z) \Big) \nn \\
& \qquad  +(y-z)^{-1}(d_{\al 1} d_{\bt 1} d_{\g 1} d_{\dt 1} e^{-2\rho -i \sg})(z) \Big] \nn \\
& \qquad= -\frac{1}{24}\ep^{\al \bt \g \dt} \Big(-i(\Pi_{\al \bt}(d_{\g1}(d_{\dt 1}e^{-\rho})))(z) + i (d_{\bt 1}(\Pi_{\al \g}(d_{\dt 1} e^{-\rho})))(z)\nn \\
& \qquad -i (d_{\bt1}(d_{\g1}(\Pi_{\al \dt}e^{-\rho})))(z) \Big) - e^{-2\rho -i \sg} (d_1)^4(z) \nn \\
&\qquad=- e^{-2\rho -i \sg} (d_1)^4(z) + \frac{i}{4}(e^{-\rho}(d_{\al 1}(d_{\bt 1} \Pi^{\al \bt})))(z) + \frac{3}{4}(e^{-\rho} d_{\al 1} \pd^2 \ta^{\al 2})(z)\,,\label{4.1}
\end{align}
where $(d_1)^4 = \frac{1}{24}\ep^{\al \bt \g \dt}d_{\al 1} d_{\bt 1} d_{\g 1} d_{\dt 1}$ and, to get the last line, we used that \\ $-i(e^{-\rho}(d_{\al 1}(\Pi^{\al \bt} d_{\bt1})))=-i(e^{-\rho}(d_{\al 1}(d_{\bt 1} \Pi^{\al \bt}))) -3(e^{-\rho} d_{\al 1} \pd^2 \ta^{\al 2})$ and \\ $-i(e^{-\rho}(\Pi^{\al \bt}(d_{\al 1} d_{\bt 1})))= -i(e^{-\rho}(d_{\al 1}(d_{\bt 1} \Pi^{\al \bt}))) -6(e^{-\rho} d_{\al 1} \pd^2 \ta^{\al 2})$.

The next terms are
\begin{align}
&-\frac{1}{24}\ep^{\al \bt \g \dt} \oint dy \, D_{\al}(y)\ep_{\ep \bt \g \dt}(\pd \ta^{\ep 1} e^{\rho + 2 i \sg})(z) \nn \\
&\qquad=\frac{1}{4} \oint dy \, (d_{\al 1} e^{-\rho -i\sg})(y) (\pd \ta^{\al 1} e^{\rho + 2 i \sg})(z)\nn \\
&\qquad= \frac{1}{4}\oint dy \oint \frac{dx}{x-z}\,\Big( 4(y-x)^{-1} \pd e^{-\rho -i \sg}(x) e^{\rho +2 i \sg}(z) \nn \\
& \qquad + \pd \ta^{\al 1}(x) (y-z)^{-1}(d_{\al 1} e^{i \sg})(z) \Big) \nn \\
&\qquad= - \frac{1}{4}(d_{\al 1}(\pd \ta^{\al 1} e^{i \sg}))(z) - \frac{1}{2}(\pd(\rho + i \sg)(\pd(\rho + i \sg)e^{i\sg}))(z) + \frac{1}{2}(\pd^2(\rho + i \sg) e^{i \sg})(z)\,.\label{4.2}
\end{align}

\begin{align}
&-\frac{1}{24}\ep^{\al \bt \g \dt} \oint dy \, D_{\al}(y)\ep_{\ep \bt \g \dt} (\pd^2 \ta^{\ep 2} e^{i \sg})(z) \nn \\
&\qquad=\frac{1}{4} \oint dy \, (d_{\al 1} e^{-\rho -i \sg})(y) (\pd^2 \ta^{\al 2} e^{i\sg})(z) \nn \\
&\qquad=\frac{1}{4}(d_{\al 1} \pd^2 \ta^{\al 2} e^{-\rho})(z)\,.\label{4.3}
\end{align}

\begin{align}
&-\frac{1}{24}\ep^{\al \bt \g \dt} \oint dy \, D_{\al}(y)\ep_{\ep \bt \g \dt}(\pd \ta^{\ep 2}(\pd(\rho + i \sg)e^{i\sg}))(z) \nn \\
&\qquad=\frac{1}{4} \oint dy \, (d_{\al 1} e^{-\rho -i \sg})(y)(\pd \ta^{\al 2}(\pd(\rho + i \sg)e^{i\sg}))(z) \nn \\
&\qquad=\frac{1}{4}(d_{\al1} (\pd \ta^{\al 2}(\pd(\rho + i \sg)e^{-\rho}))) (z)\,. \label{4.4}
\end{align}

When contracted with $-\frac{1}{24}\ep^{\al \bt \g \dt}$, the last three terms of \eqref{th} amount to \\$-\frac{i}{4}((d_{\bt 1} \Pi^{\al \bt})e^{i \sg})$. Therefore, we are left with the expression
\begingroup
\allowdisplaybreaks
\begin{align}
&-\frac{i}{4} \oint dy \, D_{\al}(y)((d_{\bt 1} \Pi^{\al \bt})e^{i \sg})(z) \nn \\
&\qquad=-\frac{i}{4} \oint dy \,(d_{\al 2} - d_{\al1} e^{-\rho - i \sg})(y)((d_{\bt 1} \Pi^{\al \bt})e^{i \sg})(z)\nn \\
&\qquad= -\frac{i}{4} \oint dy\,(d_{\al 2} - d_{\al1} e^{-\rho - i \sg})(y)\oint \frac{dx}{x-z}\,(d_{\bt 1} \Pi^{\al \bt})(x)e^{i\sg}(z)\nn\\
&\qquad=-\frac{i}{4} \oint dy \oint \frac{dx}{x-z}\, \Big[ \Big(-2i(y-x)^{-1} \Pi^m \Pi_m(x) -3i (y-x)^{-1}(d_{\al 1} \pd \ta^{\al 1})(x)\Big)\times \nn \\
& \qquad \times e^{i \sg}(z) \nn \\
& \qquad -3i (y-x)^{-1}(d_{\al 1} \pd \ta^{\al 2} e^{-\rho -i \sg})(x)e^{i\sg}(z) - (d_{\bt 1} \Pi^{\al \bt})(x) (y-z)^{-1} (e^{-\rho}d_{\al 1})(z) \Big] \nn \\
&\qquad=-\frac{1}{2}(\Pi^{\ab} \Pi_{\ab} e^{i \sg})(z)- \frac{3}{4} (d_{\al 1} \pd \ta^{\al 1}e^{i\sg})(z)- \frac{3}{4}((d_{\al 1} \pd \ta^{\al 2} e^{-\rho -i \sg})e^{i\sg})(z) \nn \\
& \qquad+ \frac{i}{4}((d_{\bt 1} \Pi^{\al \bt})(e^{-\rho}d_{\al 1}))(z) \nn \\
&\qquad= -\frac{1}{2}(\Pi^{\ab} \Pi_{\ab} e^{i \sg})(z)- \frac{3}{4} (d_{\al 1} \pd \ta^{\al 1}e^{i\sg})(z) + \frac{3}{4}(d_{\al 1}(\pd \ta^{\al 2}(\pd(\rho + i \sg)e^{-\rho})))(z) \nn \\
& \qquad+ \frac{i}{4}(e^{-\rho}(d_{\al 1}(d_{\bt 1} \Pi^{\al \bt})))(z)\,. \label{4.5}
\end{align}
\endgroup
To obtain the last line we used that $((d_{\al1} \pd \ta^{\al 2} e^{-\rho -i \sg})e^{i\sg})= (\pd(d_{\al 1} \pd \ta^{\al 2})e^{-\rho})$ \\ $- (d_{\al 1}(\pd \ta^{\al 2}(\pd(\rho + i \sg)e^{-\rho}))) $ and $((d_{\bt1} \Pi^{\al \bt})(e^{-\rho} d_{\al 1})) = (e^{-\rho}(d_{\al 1}(d_{\bt 1} \Pi^{\al \bt})))$ \\$ - 3i(\pd(d_{\al 1} \pd \ta^{\al 2})e^{-\rho})$.

Gathering eqs.~\eqref{4.1}--\eqref{4.5}, we obtain our final expression 
\begin{align}
G^+_{\rm hyb} & = - (d_1)^4 e^{-2\rho -i \sg} + \frac{i}{2}d_{\al 1}d_{\bt 1} \Pi^{\al \bt}e^{-\rho} + d_{\al1} \pd \ta^{\al 2}\pd (\rho + i \sg)e^{-\rho} +  d_{\al 1} \pd^2 \ta^{\al 2}e^{-\rho}  \nn \\
&  -\frac{1}{2}\Pi^{\ab} \Pi_{\ab} e^{i \sg} - d_{\al 1}\pd \ta^{\al 1} e^{i \sg} - \frac{1}{2}\pd ( \rho + i \sg) \pd ( \rho + i \sg) e^{i\sg}  \nn \\
&  + \frac{1}{2}\pd^2 ( \rho + i \sg) e^{i \sg} + G^+_C \,,\label{gplust}
\end{align}
where we have dropped the normal-ordering brackets.

\section{\boldmath $d=6$ $\mathcal{N}=1$ super-Yang-Mills} \label{sixsymapp}

In this section, we closely follow the $d=10$ $\mathcal{N}=1$ super-Yang-Mills description presented in ref.~\cite[Appendix B]{Mafra:2008gkx}.

To describe $d=6$ super-Yang-Mills in $\mathcal{N}=1$ superspace, we define the super-covariant derivatives
\begin{subequations}
\begin{align}
\mathfrak{D}_{\ab} & = \pd_{\ab} + A_{\ab} \,, \\
\mathfrak{D}_{\al j} & = \nabla_{\al j} + A_{\al j} \,,
\end{align}
\end{subequations}
where $\nabla_{\al j} = \frac{\pd}{\pd \ta^{\al j}} - \frac{i}{2} \ep_{jk} \ta^{\bt k}\sg^{\ab}_{\al \bt} \pd_{\ab} $ with $\{ \nabla_{\al j}, \nabla_{\bt k} \} = -i \ep_{jk} \sg^{\ab}_{\al \bt} \pd_{\ab}$. Then, the field-strengths are
\begin{subequations}
\begin{align}
F_{\al j  \bt k} & = \{ \mathfrak{D}_{\al j} , \mathfrak{D}_{\bt k} \} + i \ep_{jk} \sg^{\ab}_{\al \bt} \mathfrak{D}_{\ab} \,, \\
F_{\al j \ab} & = [\mathfrak{D}_{\al j}, \mathfrak{D}_{\ab} ] \,, \\
F_{\ab \bb} & = [\mathfrak{D}_{\ab}, \mathfrak{D}_{\bb} ] \,,
\end{align}
\end{subequations}
which are invariant under the gauge transformations
\begin{align}\label{gaugegaugeapp}
\dt A_{\al j} & = \nabla_{\al j} \La \,, & \dt A_{\ab} & = \pd_{\ab} \La \,,
\end{align}
for any $\La$.

Explicitly, the superspace field-strength constraint $F_{\al j \bt k} =0$ reads \cite{Howe:1983fr}
\begin{align}
\nabla_{\al j} A_{\bt k} + \nabla_{\bt k } A_{\al j} + \{ A_{\al j}, A_{\bt k} \} + i \ep_{jk} \sg^{\ab}_{\al \bt} A_{\ab} = 0 \,.
\end{align}
Multiplying the above equation by $(\sg^{\ab \bb \cb})^{\al \bt}$ and using that $(\sg^{\ab \bb \cb})^{\al \bt} \sg^{\db}_{\al \bt} =0$, we obtain
\begin{align}
(\sg^{\ab \bb \cb})^{\al \bt} (\nabla_{\al j} A_{\bt k} + \nabla_{\bt k } A_{\al j} + \{ A_{\al j}, A_{\bt k} \}) & = 0 \,.
\end{align}
The converse also follows.

From the Bianchi identity
\begin{align}
[ \{ \mathfrak{D}_{\al j}, \mathfrak{D}_{\bt k} \}, \mathfrak{D}_{\g l}] + [ \{ \mathfrak{D}_{\g l}, \mathfrak{D}_{\al j} \}, \mathfrak{D}_{\bt k} ] + [ \{ \mathfrak{D}_{\bt k}, \mathfrak{D}_{\g l} \}, \mathfrak{D}_{\al j} ] = 0 \,,
\end{align}
we have
\begin{align}
i \ep_{jk} \sg^{\ab}_{\al \bt} [ \mathfrak{D}_{\ab} , \mathfrak{D}_{\g l} ] + i \ep_{lj} \sg^{\ab}_{\g \al} [\mathfrak{D}_{\ab} , \mathfrak{D}_{\bt k} ] + i \ep_{kl} \sg^{\ab}_{\bt \g} [\mathfrak{D}_{\ab} , \mathfrak{D}_{\al j} ] & = 0\,,
\end{align}
which is satisfied if $F_{\al j \ab} = -i \ep_{jk} \sg _{\ab \al \bt} W^{\bt k}$ by using the Schouten identity \cite[Appendix A]{Daniel:2024kkp}. Therefore, $F_{\al j \ab} =[\mathfrak{D}_{\al j}, \mathfrak{D}_{\ab} ]= -i \ep_{jk} \sg _{\ab \al \bt} W^{\bt k}$ gives
\begin{align}
\pd_{\ab} A_{\al j} - \mathfrak{D}_{\al j} A_{\ab} -i \ep_{jk} \sg_{\ab \al \bt} W^{\bt k} & = 0 \,.
\end{align}

The Bianchi identity
\begin{align}
[ \{ \mathfrak{D}_{\al j}, \mathfrak{D}_{\bt k} \}, \mathfrak{D}_{\ab} ] + \{ [ \mathfrak{D}_{\ab} , \mathfrak{D}_{\al j} ], \mathfrak{D}_{\bt k} \} - \{ [\mathfrak{D}_{\bt k} , \mathfrak{D}_{\ab} ] , \mathfrak{D}_{\al j} \} & = 0 \,,
\end{align}
gives
\begin{align} \label{sixsym1}
 \ep_{jk} \sg^{\bb}_{\al \bt} F_{\ab \bb} + \ep_{jl} \sg_{\ab \al \g} \mathfrak{D}_{\bt k} W^{\g l} + \ep_{kl} \sg_{\ab \bt \g} \mathfrak{D}_{\al j} W^{\g l} & = 0 \,.
\end{align}
Multiplying eq.~\eqref{sixsym1} by $\sg^{\ab \al \bt}$, we obtain
\begin{align}
\ep_{jl} \mathfrak{D}_{\al k} W^{\al l} - \ep_{kl} \mathfrak{D}_{\al j} W^{\al l} & = 0 \,,
\end{align}
which imply $\mathfrak{D}_{\al j} W^{\al j} =0$. Contracting \eqref{sixsym1} with $\sg^{\ab \bt \sg}$ and $\sg^{\ab \al \sg}$, we get
\begin{subequations} 
\begin{align}
-i \ep_{jk} (\sg^{\ab \bb})^{\sg}_{\ \al} F_{\ab \bb} -4 \ep_{lk} \mathfrak{D}_{\al j} W^{\sg l} + \ep_{lk} \dt^{\sg}_{\al} \mathfrak{D}_{\bt j} W^{\bt l} + \ep_{jk} \mathfrak{D}_{\al l} W^{\sg l} & = 0 \,, \label{sixsym3}\\
i \ep_{jk} (\sg^{\ab \bb})^{\sg}_{\ \bt} F_{\ab \bb} - 4 \ep_{lk} \mathfrak{D}_{\bt j} W^{\sg l} + \ep_{lk} \dt^{\sg}_{\bt} \mathfrak{D}_{\al j} W^{\al l} + 3 \ep_{jk} \mathfrak{D}_{\bt l} W^{\sg l} & = 0 \,. \label{sixsym4}
\end{align}
\end{subequations}
From $\eqref{sixsym4} - 3 \times \eqref{sixsym3} $, it follows that
\begin{align}
2 i \ep_{jk} (\sg^{\ab \bb})^{\bt}_{\ \al} F_{\ab \bb} - 4 \mathfrak{D}_{\al j} W^{\bt}_k + \dt^{\bt}_{\al} \mathfrak{D}_{\g j} W^{\g}_k & = 0 \,,
\end{align}
where $W^{\al}_j = \ep_{jk} W^{\al k}$. Consequently,
\begin{subequations}
\begin{align}
i \ep_{jk} (\sg^{\ab \bb})^{\bt}_{\ \al} F_{\ab \bb} - \mathfrak{D}_{\al [j}W^{\bt}_{k]}& = 0\,, \\
4 \mathfrak{D}_{\al (j} W^{\bt}_{k)} - \dt^{\bt}_{\al} \mathfrak{D}_{\g (j} W^{\g}_{k)} & = 0 \,.
\end{align}
\end{subequations}
Furthermore, using the equation of motion of $d=6$ super-Yang-Mills \cite{Howe:1983fr}, i.e., $\mathfrak{D}_{\al (j} W^{\al}_{k)} =0$, we then have
\begin{align} \label{sixsym6}
\mathfrak{D}_{\al j} W^{\bt}_k -\frac{i}{2} \ep_{jk} (\sg^{\ab \bb})^{\bt}_{\ \al} F_{\ab \bb} & =0 \,,
\end{align}

Now, consider the Bianchi identity
\begin{align}
[[\mathfrak{D}_{\ab}, \mathfrak{D}_{\bb}], \mathfrak{D}_{\al j} ] + [[ \mathfrak{D}_{\al j}, \mathfrak{D}_{\ab}], \mathfrak{D}_{\bb}] + [[ \mathfrak{D}_{\bb}, \mathfrak{D}_{\al j} ], \mathfrak{D}_{\ab} ]& = 0\,,
\end{align}
that implies
\begin{align} \label{sixsym5}
\mathfrak{D}_{\al j} F_{\ab \bb} & = i \ep_{jk} \sg_{\ab \al \bt} \mathfrak{D}_{\bb} W^{\bt k} - i \ep_{jk} \sg_{\bb \al \bt} \mathfrak{D}_{\ab} W^{\bt k} \,.
\end{align}
Finally, acting with $\mathfrak{D}_{\g l}$ in \eqref{sixsym6}, symmetrizing in the indices $\{ \al j , \g l \}$, then using \eqref{sixsym5} and multiplying by $\dt^{\g}_{\bt}$, we end up with
\begin{align}
\sg^{\ab}_{\al \bt} \mathfrak{D}_{\ab} W^{\bt j} & = 0 \,.
\end{align}

In summary, the equations describing $d=6$ super-Yang-Mills obtained in this section are 
\begin{subequations} \label{sixsym7}
\begin{align}
\nabla_{\al j} A_{\bt k} + \nabla_{\bt k } A_{\al j} + \{ A_{\al j}, A_{\bt k} \} + i \ep_{jk} \sg^{\ab}_{\al \bt} A_{\ab} & = 0  \,, \\
\pd_{\ab} A_{\al j} - \mathfrak{D}_{\al j} A_{\ab} -i \ep_{jk} \sg_{\ab \al \bt} W^{\bt k} & = 0  \,, \\
 \mathfrak{D}_{\al j} W^{\bt}_k -\frac{i}{2} \ep_{jk} (\sg^{\ab \bb})^{\bt}_{\ \al} F_{\ab \bb} & = 0 \,, \label{sixsym8} \\
 \sg^{\ab}_{\al \bt} \mathfrak{D}_{\ab} W^{\bt j} & = 0  \,, \label{sixsym9}
\end{align}
\end{subequations}
which were shown to follow from the equation of motion $ \mathfrak{D}_{\al (j} W^{\al}_{k)} = 0$ and the superspace constraint $F_{\al j \bt k} =0$. 

Note also that the superfields $ \{ A_{\ab}, W^{\al j}, F_{\ab \bb} \}$ can be written as
\begin{subequations}
\begin{align}
A_{\ab} & = - \frac{i}{4} \ep^{jk} \sg_{\ab}^{\al \bt} ( \nabla_{\al j} A_{\bt k} + \nabla_{\bt k} A_{\al j} + \{ A_{\al j}, A_{\bt k} \} ) \,, \label{gaugefieldsix}\\
W^{\al j} & = \frac{i}{3} \ep^{jk} \sg^{\ab \al \bt}  ( \pd_{\ab} A_{\bt k}- \mathfrak{D}_{\bt k} A_{\ab} ) \,, \\
F_{\ab \bb} & = \mathfrak{D}_{\ab} A_{\bb} - \mathfrak{D}_{\bb} A_{\ab}\,.
\end{align}
\end{subequations}

 The $\ta$ expansion of the $d=6$ SYM superfields is given by
 \begin{subequations} \label{thetaexpSYM}
\begin{align}
A_{\al j } & = - \frac{i}{2} \ep_{jk} a_{\al \bt} \ta^{\bt k} + \frac{1}{3} \ep_{\al \bt \g \dt} \ep_{jk} \ep_{lm} \ta^{\bt k} \psi^{\g l} \ta^{\dt m} + \ldots \,, \\
A_{\ab} & = a_{\ab} + i \ep_{jk} \sg_{\ab \al \bt} \psi^{\al j} \ta^{\bt k} + \ldots \,, \\
W^{\al j} & = \psi^{\al j} - \frac{i}{2} (\sg^{\ab \bb})^{\al}_{\ \bt} \ta^{\bt j} f_{\ab \bb} + \ldots \,, \\
F_{\ab \bb} & = f_{\ab \bb} + \ldots \,,
\end{align}
\end{subequations}
where $a_{\ab}$ is the gluon, $\psi^{\al j}$ the gluino and $f_{\ab \bb} = \pd_{\ab} a_{\bb} - \pd_{\bb} a_{\ab}$ the gluon field-strength. Note further that the first component of $A_{\al j}$ can be gauged away.

\section{\boldmath BRST invariance of the $ d=6$ $\mathcal{N}=1$ supersymmetric vertex operator $U$} \label{newvertexapp}

In this section, we prove that BRST invariance of the vertex operator \eqref{gh1vertex} implies the $d=6$ SYM equations of motion in superspace 
\begin{subequations}\label{newvertexappeqs}
\begin{align}
(\sg^{\ab \bb \cb})^{\al \bt} (\nabla_{\al j} A_{\bt k} + \nabla_{\bt k} A_{\al j})&  = 0 \,, \label{quantapp01}\\
\nabla_{\al j} W^{\bt k} + \frac{i}{2} \dt^k_j (\sg_{\ab \bb})^{\bt}_{\ \al} F^{\ab \bb} & = 0 \label{quantapp02}\,, 
\end{align}
\end{subequations}
where, as shown in Appendix \ref{sixsymapp}, the remaining $d=6$ $\mathcal{N}=1$ superfields are defined in terms of $A_{\al j}$ according to
\begin{subequations}\label{newvertexappeqs2}
\begin{align}
A_{\ab} & = - \frac{i}{4} \ep^{jk} \sg_{\ab}^{\al \bt} ( \nabla_{\al j} A_{\bt k} + \nabla_{\bt k} A_{\al j} ) \,, \\
W^{\al j} & = \frac{i}{3} \ep^{jk} \sg^{\ab \al \bt}  ( \pd_{\ab} A_{\bt k}- \nabla_{\bt k} A_{\ab} ) \,, \\
F_{\ab \bb} & = \pd_{\ab} A_{\bb} - \pd_{\bb} A_{\ab}\,.
\end{align}
\end{subequations}

The BRST operator is given by the zero-mode of the superconformal generator $G^+$ defined in \eqref{NMGplus}. For the purpose of showing that $(G^+)_0 U=0$ implies eqs. \eqref{newvertexappeqs}, it is convenient to systematically divide up the calculation into linearly independent terms, each characterized by a specific power of the ghost fields. In addition, since the vertex operator $U$ is annihilated by $(G^-)_0$, recall that it satisfies the Lorenz gauge condition, which implies that the vector superfield is divergenceless $\pd^{\ab} A_{\ab} = 0$.

Before providing the computational details of the vertex operator BRST invariance, we recall the following identities, which can be derived from the Lorenz gauge condition as well as from eqs.~\eqref{newvertexappeqs} and \eqref{newvertexappeqs2}
\begin{subequations}
\begin{align}
\nabla_{\al 1} A_{\bt 2} + \nabla_{\bt 2} A_{\al 1} + \nabla_{\bt 1} A_{\al 2} + \nabla_{\al 2} A_{\bt 1} & = 0 \,, \\
A_{\ab} - \frac{i}{2} \sg_{\ab}^{\g \dt} \big( \nabla_{\g 1} A_{\dt 2} + \nabla_{\dt 2} A_{\g 1} \big) & = 0 \,, \\
\sg_{\ab \al \dt} A^{\bt \dt} + \frac{i}{2} \sg_{\ab}^{\dt \bt} \big( \nabla_{\dt 1} A_{\al 2} + \nabla_{\al 2} A_{\dt 1} \big) - \frac{i}{2} \sg_{\ab}^{\dt \bt} \big( \nabla_{\al 1} A_{\dt 2} + \nabla_{\dt 2} A_{\al 1} \big) & \nn \\
- \frac{i}{2} \dt^{\al}_{\bt} \sg_{\ab}^{\g \dt} \big( \nabla_{\g 1} A_{\dt 2} + \nabla_{\dt 2} A_{\g 1} \big) & = 0 \,, \\
\pd_{\ab} A_{\al j} - \nabla_{\al j} A_{\ab} - i \ep_{jk} \sg_{\ab \al \bt} W^{\bt k} & = 0 \,.
\end{align}
\end{subequations}
In addition, we also have the equation of motion
\begin{align}
\nabla_{\al 2} W^{\bt 2} - i \pd^{\bt \g} \big( \nabla_{\g 1} A_{\al 2} + \nabla_{\al 2} A_{\g 1} \big) & = 0 \,,
\end{align}
and note that from
\begin{align}
\nabla_{\al 1} A_{\ab} & = \frac{i}{2} \sg_{\ab}^{\bt \g} \nabla_{\bt 1} \nabla_{\g 1} A_{\al 2} - \sg_{\ab}^{\bt \g} \pd_{\al \bt} A_{\g 1} + \pd_{\ab} A_{\al 1} - \frac{i}{2} \sg_{\ab}^{\bt \g} \nabla_{\al 2} \nabla_{\bt 1} A_{\g 1} \,,
\end{align}
the spinor field-strength Bianchi identity can be written as
\begin{align}\label{bianchiappfinance}
0 & = i \sg_{\ab \al \bt} W^{\bt 2} - \pd_{\ab} A_{\al 1} + \nabla_{\al 1} A_{\ab} \nn \\
& = i \sg_{\ab \al \bt} W^{\bt 2} + \frac{i}{2} \sg_{\ab}^{\bt \g} \nabla_{\bt 1} \nabla_{\g 1} A_{\al 2} - \sg_{\ab}^{\bt \g} \pd_{\al \bt} A_{\g 1} - \frac{i}{2} \sg_{\ab}^{\bt \g} \nabla_{\al 2} \nabla_{\bt 1} A_{\g 1} \,.
\end{align}

In what follows, we outline the main steps arising from the action of $(G^+)_0$ in the unintegrated vertex $U$, where the calculation is structured by distinct powers of the ghost fields.

\paragraph{Terms proportional to $\la^{\al} \la^{\bt}$.} The terms proportional to $\la^{\al} \la^{\bt}$ in the operator $(G^+)_0 U$ are given by
\begin{align}
\la^{\al} \la^{\bt} u^j u^k \nabla_{\al j} A_{\bt k}\,, \label{quantapp1}
\end{align}
where we are defining $u^j = \{-e^{-\rho -i \sg}, 1\}$ and, as usual, $\nabla_{\al j}$ is the zero-mode of $d_{\al j}$. Therefore, the vanishing of eq.~\eqref{quantapp1} implies the constraint \eqref{quantapp01} on the $d=6$ spinor superfield $A_{\al j}$.

\paragraph{Terms proportional to $\la^{\al} e^{i\sg}$.} From $(G^+)_0 U$, one finds that the expression with terms proportional to $\la^{\al} e^{i\sg}$ is given by
\begin{align}
& \la^{\al} \bigg[ \pd \ta^{\bt 1} \bigg( \nabla_{\bt 1} A_{\al 2} + \nabla_{\al 2} A_{\bt 1} - i A_{\al \bt} \bigg) + \Pi^{\ab} \bigg( \pd_{\ab} A_{\al 2} - \nabla_{\al 2} A_{\ab} + i \sg_{\ab \al \bt} W^{\bt 1} \bigg) \nn \\
& + d_{\bt 1} \nabla_{\al 2} W^{\bt 1} \bigg] e^{i \sg} \,, \label{secondeqneq1}
\end{align}
which vanishes by the SYM equations of motion \eqref{newvertexappeqs} and the superspace definitions \eqref{newvertexappeqs2}.

\paragraph{Terms proportional to $\la^{\al} e^{-\rho}$.} From $(G^+)_0 U$, one finds that
\begin{align}
& \la^{\al} \bigg[ d_{\bt 1} \Pi^{\ab} \bigg( \dt^{\bt}_{\al} A_{\ab} - \sg_{\ab \al \g} A^{\bt \g} + i \sg_{\ab}^{\bt \g} \nabla_{\g 1} A_{\al 2} + i \sg_{\ab}^{\bt \g} \nabla_{\al 2} A_{\g 1} \bigg) \nn \\
& + \frac{i}{2} \sg^{\ab \g \dt} d_{\g 1} d_{\dt 1} \bigg( - i \sg_{\ab \al \bt} W^{\bt 1} - \pd_{\ab} A_{\al 2} + \nabla_{\al 2} A_{\ab} \bigg)  + \pd^2 \ta^{\bt 2} \bigg( i A_{\al \bt}\nn \\
& - \nabla_{\al 1} A_{\al 2} - \nabla_{\al 2} A_{\bt 1} \bigg) + \pd \Pi^{\ab} \bigg( \pd_{\ab} A_{\al 1} + i \sg_{\ab \al \bt} W^{\bt 2} + \sg_{\ab \al \bt} \pd^{\bt \g} A_{\g 1} \nn \\
& + \frac{i}{2} \sg_{\ab}^{\bt \g} \nabla_{\bt 1} \nabla_{\g 1} A_{\al 2} - \frac{i}{2} \sg_{\ab}^{\bt \g} \nabla_{\al 2} \nabla_{\bt 1} A_{\g 1} \bigg) + \frac{1}{2} \pd^2 \ta^{\bt 1} \bigg( \nabla_{\bt 1} A_{\al 1} + \nabla_{\al 1} A_{\bt 1} \bigg) \nn \\
& + \pd d_{\bt 1} \bigg( \nabla_{\al 2} W^{\bt 2} - i \pd^{\bt \g} \nabla_{\al 2} A_{\g 1} - i \pd^{\bt \g} \nabla_{\g 1} A_{\al 2} \bigg) \bigg]\,, \label{secondeqnew2}
\end{align}
which again vanishes by using eqs.~\eqref{newvertexappeqs} and \eqref{newvertexappeqs2}.

\paragraph{Terms proportional to $\la^{\al}\pd \rho e^{-\rho}$ and $\la^{\al} i\pd\sg e^{-\rho}$.} From $(G^+)_0 U$, one finds that
\begin{align}
&\la^{\al} \bigg\{ \bigg[ \frac{1}{2} \pd \ta^{\bt 1}  \bigg( \nabla_{\al 1} A_{\bt 1} + \nabla_{\bt 1} A_{\al 1} \bigg) + \pd \ta^{\bt 2} \bigg( - \nabla_{\al 2} A_{\bt 1} - \nabla_{\bt 1} A_{\al 2} - i A_{\al \bt} \bigg) \nn \\
& + \Pi^{\ab} \bigg( - \nabla_{\al 1} A_{\ab} - \frac{i}{2} \sg_{\ab}^{\bt \g} \nabla_{\bt 1} \nabla_{\g 1} A_{\al 2} -2i\sg_{\ab \al \bt} W^{\bt 2} - \sg_{\ab \al \bt} \pd^{\bt \g} A_{\g 1} + \frac{i}{2} \sg_{\ab}^{\bt \g} \nabla_{\al 2} \nabla_{\bt 1} A_{\g 1} \bigg) \nn \\
& + d_{\bt 1} \bigg( \nabla_{\al 1} W^{\bt 1} +i \pd^{\bt \g} \nabla_{\g 1} A_{\al 2} - 2 \nabla_{\al 2} W^{\bt 2} + i \pd^{\bt \g} \nabla_{\al 2} A_{\g 1} \bigg)\bigg] \pd \rho e^{-\rho} + \bigg[ \pd \ta^{\bt 1} \bigg( \nabla_{\al 1} A_{\bt 1} \nn\\ 
& + \nabla_{\bt 1} A_{\al 1} \bigg)  + \pd \ta^{\bt 2} \bigg( - \nabla_{\al 2} A_{\bt 1} - \nabla_{\bt 1} A_{\al 2} - i A_{\al \bt} \bigg) + \Pi^{\ab} \bigg( \pd_{\ab} A_{\al 1} - \nabla_{\al 1} A_{\ab} \nn \\
& - i \sg_{\ab \al \bt} W^{\bt 2} \bigg) + d_{\bt 1} \bigg( \nabla_{\al 1} W^{\bt 1} - \nabla_{\al 2} W^{\bt 2} \bigg) \bigg] i \pd \sg e^{-\rho} \,, \label{secondeqnew3}
\end{align}
which vanishes by using eqs.~\eqref{newvertexappeqs} and \eqref{newvertexappeqs2}.

Note that from $\nabla_{\al 2} W^{\bt 2} - i \pd^{\bt \g} \big( \nabla_{\g 1} A_{\al 2} + \nabla_{\al 2} A_{\g 1} \big) = 0 $, we have $i \pd^{\bt \g} \nabla_{\g 1} A_{\al 2} = \nabla_{\al 2} W^{\bt 2} - i \pd^{\bt \g} \nabla_{\al 2} A_{\g 1}$, so that
\begin{align}
\nabla_{\al 1} W^{\bt 1} + i \pd^{\bt \g} \nabla_{\g 1} A_{\al 2} - 2 \nabla_{\al 2} W^{\bt 2} + i \pd^{\bt \g} \nabla_{\al 2} A_{\g 1} & = \nabla_{\al 1}W^{\bt 1} - \nabla_{\al 2} W^{\bt 2}\,,
\end{align}
which can be seen to vanish by symmetrizing the $\rm SU(2)$ indices of eq.~\eqref{quantapp02}.

Also, from 
\begin{align}
\nabla_{\al 1} A_{\ab} & = \frac{i}{2} \sg_{\ab}^{\bt \g} \nabla_{\bt 1} \nabla_{\g 1} A_{\al 2} - \sg_{\ab}^{\bt \g} \pd_{\al \bt} A_{\g 1} + \pd_{\ab} A_{\al 1} - \frac{i}{2} \sg_{\ab}^{\bt \g} \nabla_{\al 2} \nabla_{\bt 1} A_{\g 1} \,,
\end{align}
and
\begin{align}
0 & = -2 i \sg_{\ab \al \bt} W^{\bt 2} - i \sg_{\ab}^{\bt \g} \nabla_{\bt 1} \nabla_{\g 1} A_{\al 2} + 2 \sg_{\ab}^{\bt \g} \pd_{\al \bt} A_{\g 1} + i \sg_{\ab}^{\bt \g} \nabla_{\al 2} \nabla_{\bt 1} A_{\g 1} \,,
\end{align}
we have that
\begin{align}
 - \nabla_{\al 1} A_{\ab} - \frac{i}{2} \sg_{\ab}^{\bt \g} \nabla_{\bt 1} \nabla_{\g 1} A_{\al 2} -2i\sg_{\ab \al \bt} W^{\bt 2} - \sg_{\ab \al \bt} \pd^{\bt \g} A_{\g 1} + \frac{i}{2} \sg_{\ab}^{\bt \g} \nabla_{\al 2} \nabla_{\bt 1} A_{\g 1} & = 0\,.
\end{align}

\paragraph{Terms proportional to $\la^{\al} (d_1^3)^{\bt}e^{-2 \rho -i \sg}$.} This contribution takes the form
\begin{align}
\la^{\al} (d_1^3)^{\bt} \bigg( \nabla_{\al 2} A_{\bt 1} + \nabla_{\bt 1} A_{\al 2} - i A_{\al \bt} \bigg) e^{-2 \rho -i \sg} \,,
\end{align}
and it vanishes by the definition of the vector connection.

\paragraph{Terms proportional to $\la^{\al} d_{\bt 1} \pd d_{\g 1}e^{-2 \rho -i \sg}$.} This contribution vanishes by the Bianchi identity \eqref{bianchiappfinance}.

\paragraph{Terms proportional to $\la^{\al} \pd^2 d_{\bt 1}e^{-2 \rho -i \sg}$.} This contribution gives
\begin{align}
&\frac{1}{2} \la^{\al} \pd^2 d_{\bt 1} \bigg( -(\nabla_1^3)^{\bt} A_{\al 2} + i \pd^{\bt \g} \nabla_{\g 1} A_{\al 1} + \frac{i}{2} \dt^{\bt}_{\al} \pd^{\g \dt} \nabla_{\g 1} A_{\dt 1}\nn \\
& - \frac{1}{6} \ep^{\bt \g \dt \sg} \nabla_{\al 2} \nabla_{\g 1} \nabla_{\dt 1} A_{\sg 1} \bigg) e^{-2 \rho -i \sg} \,,\label{secondeqnew4}
\end{align}
which vanishes by eq.~\eqref{quantapp02}.

The terms proportional to $\la^{\al}d_{\bt 1} d_{\g 1} \Pi^{\ab}e^{-2 \rho -i \sg}$, $\la^{\al} \pd d_{\bt 1} \Pi^{\ab}e^{-2 \rho -i \sg}$, $\la^{\al}  d_{\bt 1} \pd \Pi^{\ab}e^{-2 \rho -i \sg}$, $\la^{\al} \pd^2 \ta^{\bt 2} e^{-2 \rho -i \sg}$, $\la^{\al} \pd^2 \Pi^{\ab} e^{-2 \rho -i \sg}$, $\la^{\al} \pd^3 \ta^{\bt 2} e^{-2 \rho -i \sg}$, $\la^{\al} d_{\bt 1} d_{\g 1} \pd e^{-2 \rho -i \sg}$, $\la^{\al} \pd d_{\bt 1} \pd e^{-2 \rho -i \sg}$, $\la^{\al} d_{\bt 1} \Pi^{\ab} \pd e^{-2 \rho -i \sg}$, $\la^{\al} \pd^2 \ta^{\bt 2} \pd e^{-2 \rho -i \sg}$, $\la^{\al} \pd \Pi^{\ab} \pd e^{-2 \rho -i \sg}$, $\la^{\al} d_{\bt 1} \pd^2 e^{-2 \rho -i \sg}$, $\la^{\al} \Pi^{\ab} \pd^2e^{-2 \rho -i \sg}$, $\la^{\al} \pd \ta^{\bt 2} \pd^2 e^{-2 \rho -i \sg}$ and $\la^{\al} \pd^3 e^{-2 \rho -i \sg}$ can be similarly shown to yield a vanishing result.

Concerning the supersymmetric $d=6$ SYM equations of motion \eqref{newvertexappeqs}, it is important to note that the expression \eqref{quantapp1} gives rise to the off-shell constraint in eq.~\eqref{quantapp01}, and the contributions \eqref{secondeqneq1}, \eqref{secondeqnew2}, \eqref{secondeqnew3} and \eqref{secondeqnew4} imply the $d=6$ SYM equation of motion in superspace, namely, eq.~\eqref{quantapp02}.

Finally, we should mention that since the vertex operator is annihilated by the BRST operator $(G^+)_0$ and by $(G^-)_0$ (Lorenz gauge condition), it is also annihilated by $T_0$, i.e., it is a primary field. As an independent check, we have also confirmed that $T_0 U=0$, which gives further support for our supersymmetric vertex operator description. 




\bibliography{bibliomain}
\bibliographystyle{JHEP.bst}

\end{document}